\documentclass[%
 aip,
 amsmath,amssymb,
floatfix,%
reprint,%
]{revtex4-1}

\usepackage{graphicx}
\usepackage{dcolumn}
\usepackage{bm}
\usepackage{xcolor}
\usepackage[normalem]{ulem}
\usepackage{tabularx}

\usepackage[utf8]{inputenc}
\usepackage[T1]{fontenc}
\usepackage{mathptmx}
\usepackage{etoolbox}
\usepackage[shortlabels]{enumitem}

\usepackage{booktabs}

\newcommand{\fref}[1]{Fig.~\ref{#1}}
\newcommand{\tref}[1]{Table~\ref{#1}}
\newcommand{\eref}[1]{Eq.~\ref{#1}}
\newcommand{\cref}[1]{Chapter~\ref{#1}}
\newcommand{\sref}[1]{Sec.~\ref{#1}}
\newcommand{\aref}[1]{Appendix~\ref{#1}}

\makeatletter
\def\@email#1#2{%
 \endgroup
 \patchcmd{\titleblock@produce}
  {\frontmatter@RRAPformat}
  {\frontmatter@RRAPformat{\produce@RRAP{*#1\href{mailto:#2}{#2}}}\frontmatter@RRAPformat}
  {}{}
}%
\makeatother
\begin{document}

\preprint{AIP/123-QED}

\title[CaBER polymer solutions]{Numerical analysis of capillarity-driven thinning rheometry for polydisperse polymer solutions}
\author{Isaac Pincus}
 \affiliation{Department of Mechanical Engineering, Massachusetts Institute of Technology, Cambridge, Massachusetts, USA}
 \affiliation{Institute of Earth Sciences, University of Lausanne, Lausanne, Switzerland}
 
\author{Vincenzo Calabrese}%
\affiliation{ 
Okinawa Institute of Science and Technology Graduate University, Onna-son, Okinawa, Japan
}%
\affiliation{ POLYMAT, Rheology and Advanced Manufacturing group, University of the Basque Country UPV/EHU, Donostia-San Sebastian 20018, Spain}

\author{Simon J. Haward}%
\affiliation{ 
Okinawa Institute of Science and Technology Graduate University, Onna-son, Okinawa, Japan
}%

\author{Gareth H. McKinley}
\affiliation{Department of Mechanical Engineering, Massachusetts Institute of Technology, Cambridge, Massachusetts, USA}

\date{\today}

\begin{abstract}

Liquid bridges of polymer solutions that are self-thinning due to the action of capillarity undergo a transition from Newtonian-like linear thinning to exponential elastocapillary (EC) thinning when the polymer chains are stretched by the elongational flow and the resulting elastic contribution to the stress exceeds the viscous stress. As the Oldroyd-B model predicts that the EC thinning rate is set by the relaxation time ($\tau$) of the polymer, the characteristic thinning timescale extracted from the exponential decay ($\tau_{EC}$) is commonly interpreted as a direct measure of $\tau$. Here we show that for real polydisperse polymer solutions, $\tau_{EC}$ reflects only a subset of the molecular weight (MW) distribution -- those chains actively stretched by the flow. We demonstrate this using a multi-mode FENE-PM model that explicitly incorporates the molecular weight distribution, validated against the filament thinning experiments of  Calabrese et al. [Phys. Rev. X 15, 021025 (2025)] on bidisperse blends of narrowly-distributed low-MW and high-MW polystyrene solutions. The model predicts that only chains with effective Weissenberg number $Wi =  \dot{\varepsilon} \tau > 1/2 $ are extended by the flow and contribute elastic stress; this threshold naturally favors high molecular weight species, whose longer relaxation times allow them to remain stretched throughout the elastocapillary regime. The measured $\tau_{EC}$ is therefore set by this stress-contributing sub-ensemble rather than the full distribution. Further, our model predicts that $\tau_{EC}$ depends on both the molecular weight distribution and total polymer concentration, as well as experimental parameters including pre-stretch and initial filament diameter, confirming that it is best understood as an experiment-specific quantity rather than an intrinsic fluid property.

\end{abstract}

\maketitle

\section{Introduction}
\label{Introduction} 

When a polymeric fluid filament thins under the influence of capillary pressure, the resulting elongational flow within the filament induces elastic stresses that oppose further thinning.
This stress balance results in a stable cylindrical thread which resists breakup into droplets.
By measuring the evolution in the filament radius over time for a fluid with known surface tension, one can determine both the tensile stress difference and the extensional strain rate experienced by the fluid \cite{mckinley2005visco}, and hence calculate the temporal evolution of the transient extensional viscosity.
This forms the basis of experimental techniques such as the Capillary Breakup Extensional Rheometer (CaBER) \cite{bazilevsky1990liquid, anna2001elasto}.

The simplest fluid rheology which captures these elastic stresses is the Oldroyd-B constitutive model, wherein the fluid is characterised by a solvent viscosity $\eta_s$, linear elastic modulus $g$, and stress relaxation timescale $\tau$. 
For a cylindrical Oldroyd-B-fluid filament of radius $a$, through a balance of elastic, viscous, and capillary forces one can show that in the long-time limit ($t/\tau \gg 1$)\cite{bousfield1986nonlinear}:
\begin{equation}
\label{eq: filament decay rate}
    \dot{\varepsilon} = - \frac{2}{a} \frac{\mathrm{d}a}{\mathrm{d}t} \overset{t/\tau \gg 1}{=} \frac{2}{3 \tau}  \implies a \propto \exp\left(-\frac{t}{3\tau}\right)
\end{equation}
where $\dot{\varepsilon}$ is the filament decay rate (kinematically equivalent to the elongation rate experienced by a fluid element in the thinning thread), often expressed as a dimensionless Weissenberg number $W\!i = \dot{\varepsilon} \tau\overset{t/\tau \gg 1}{=} 2/3$.
As $\dot{\varepsilon}$ is a constant at $t/\tau \gg 1$, the filament is predicted to decay exponentially at a rate inversely proportional to the fluid relaxation time.
In practice, real polymeric liquids undergoing capillarity-driven self-thinning typically exhibit multiple regimes: an initial inertio-visco-capillary (IVC) regime characterized by power-law thinning dynamics in time, followed by an elastocapillary (EC) regime characterised by an exponential thinning rate as polymeric elastic stresses develop, and finally a terminal regime again characterised by a linear rate of thinning which arises from the finite extensibility of the dissolved polymer chains and ultimately leads to finite time breakup of the fluid thread \cite{entov1997effect, anna2001elasto}. 
This terminal regime is commonly modeled using a FENE (Finitely Extensible Nonlinear Elastic) extension of the Oldroyd-B fluid, incorporating a nonlinear force-extension law for the polymer chains \cite{mckinley2005visco, oliveira2006iterated, clasen2006dilute}.

While theory and simulations do not necessarily capture the complete filament behaviour, the correspondence between the Oldroyd-B prediction and the observed exponential thinning during the EC regime has led to the elastocapillary filament decay rate $\dot{\varepsilon}_\mathrm{EC}$ being used to define a characteristic `elongational relaxation time' $\tau_\mathrm{EC}$ of the polymer solution:
\begin{equation}
\label{eq: definition of tau_EC}
    \tau_\mathrm{EC} := \frac{2}{3\dot{\varepsilon}_\mathrm{EC}}
\end{equation}
where $\dot{\varepsilon}_\mathrm{EC}$ is the filament exponential decay rate during the EC thinning regime (as in \eref{eq: filament decay rate}).
However, this identification has been subject to recent debate \cite{gaillard2024beware, gaillard2025does, aisling2024importance, hu2025revealing, calabrese2024polymers}, as in many cases it does not appear to match the relaxation time expected from steady shear or linear viscoelastic measurements \cite{clasen2006dilute, liang1994rheological, calabrese2024polymers, dinic2015extensional, dinic2019macromolecular, tirtaatmadja2006drop}.
One should note that the Oldroyd-B model contains only a single relaxation time $\tau_1$, and so can not predict a different measured value in shear and extension.

A significant additional challenge arises when characterising real polymer solutions formulated from commercial material sources with broad molecular weight distributions.
This results in viscoelastic fluids with multi-modal behavior that cannot be described by a single relaxation time.
Even for ideal monodisperse polymeric systems, the chain dynamics in solution result in a well-defined Rouse-Zimm spectrum of time constants and the question naturally arises as to which moment of this distribution is measured in shear and extensional flow.
In a seminal paper, Entov and Hinch demonstrated \cite{entov1997effect} that for a dilute solution of highly extensible chains modelled as a suspension of non-interacting FENE dumbbells at times $t \gg \tau$, the longest relaxation time of the ensemble eventually sets the exponential thinning rate, which then corresponds to the measured $\tau_\mathrm{EC}$ in a CaBER experiment.
This finding relies upon a continuous decrease in the filament decay rate $\dot{\varepsilon}$ with time until \eref{eq: filament decay rate} holds with $\tau$ the \textit{longest} relaxation time of the polymer solution, with the elastic stresses growing continuously and exponentially in time to exactly balance the filament thinning rate \cite{clasen2006dilute, clasen2006beads, gaillard2025does, gaillard2024beware, hu2025revealing}.
For real polymers, however, finite extensibility effects create a plateau in the extensional viscosity at high strains, causing the thinning rate to ultimately increase again at very high stretches and eventually result in filament breakup \cite{entov1997effect}.
This increasing rate of extension will begin to stretch shorter chains in the underlying distribution, and therefore it is not necessarily true that from observing a macroscopic variable such as the rate of change of filament radius, one measures a single `longest relaxation time', but rather some average of the thinning dynamics arising from the range of polymeric species contributing to the total elastic stress in the fluid filament \cite{plog2005influence, clasen2006dilute}.
Given this ambiguity, a comprehensive framework for analyzing CaBER experiments on polydisperse fluids would be helpful for future studies and for resolving questions about the experimental insights obtained from filament thinning \cite{hu2025revealing}.

This issue has been previously investigated in the context of electrospun polymer fibers, where remarkably low concentrations of high molecular weight polymers in polydisperse solutions were found to disproportionately influence fiber formation \cite{palangetic2014dispersity, merchiers2022extensibility}. 
Long flexible polymer chains are highly extensible and generate substantial elastic stresses at small strains while maintaining a dominantly Hookean elastic response under large Hencky strains, resulting in highly stable filaments that are more resistant to droplet formation and breakup. 
Similar stabilizing effects have also been shown in bidisperse cellulose systems \cite{guizani2021fast}.
It has been demonstrated that the `spinnability' of polymer solutions—their tendency to form fibers—correlates strongly with the extensibility-weighted molecular weight $M_L$ \cite{palangetic2014dispersity}:
\begin{equation}
\label{eq: ext-weighted MW}
    M_L = \left( \sum_i w_i M_i^{\left(1 + \nu\right)} \right)^{1/(\nu + 1)}
\end{equation}
where $w_i$ represents the weight-fraction of species $i$, $M_i$ is the molecular weight of species $i$, and $\nu$ is the excluded volume exponent \cite{rubinstein2003polymer}.
This mean measure $M_L$ is derived through weighting $w_i$ by the contribution of a polymer chain with molar mass $M$ to the polymeric tensile stress difference when fully extended, which is known to scale with $M^{\left(1 + \nu\right)}$ \cite{palangetic2014dispersity}.
However, it is unclear to what extent this is the `correct' averaging of the molecular weight distribution in a transient elastocapillary regime.
During the EC regime, the local deformation rate $\dot{\varepsilon}$ is history dependent, and at a given time $t$ not all species are feeling a `strong' flow (specifically $W\!i \equiv \dot{\varepsilon}(t) \tau_i > 0.5$ above the so-called coil-stretch transition \cite{de1974coil, prabhakar2016influence}) which can extend the chains. 

\begin{figure}[t]
    \centering
    \includegraphics[width=8.5cm,height=!]{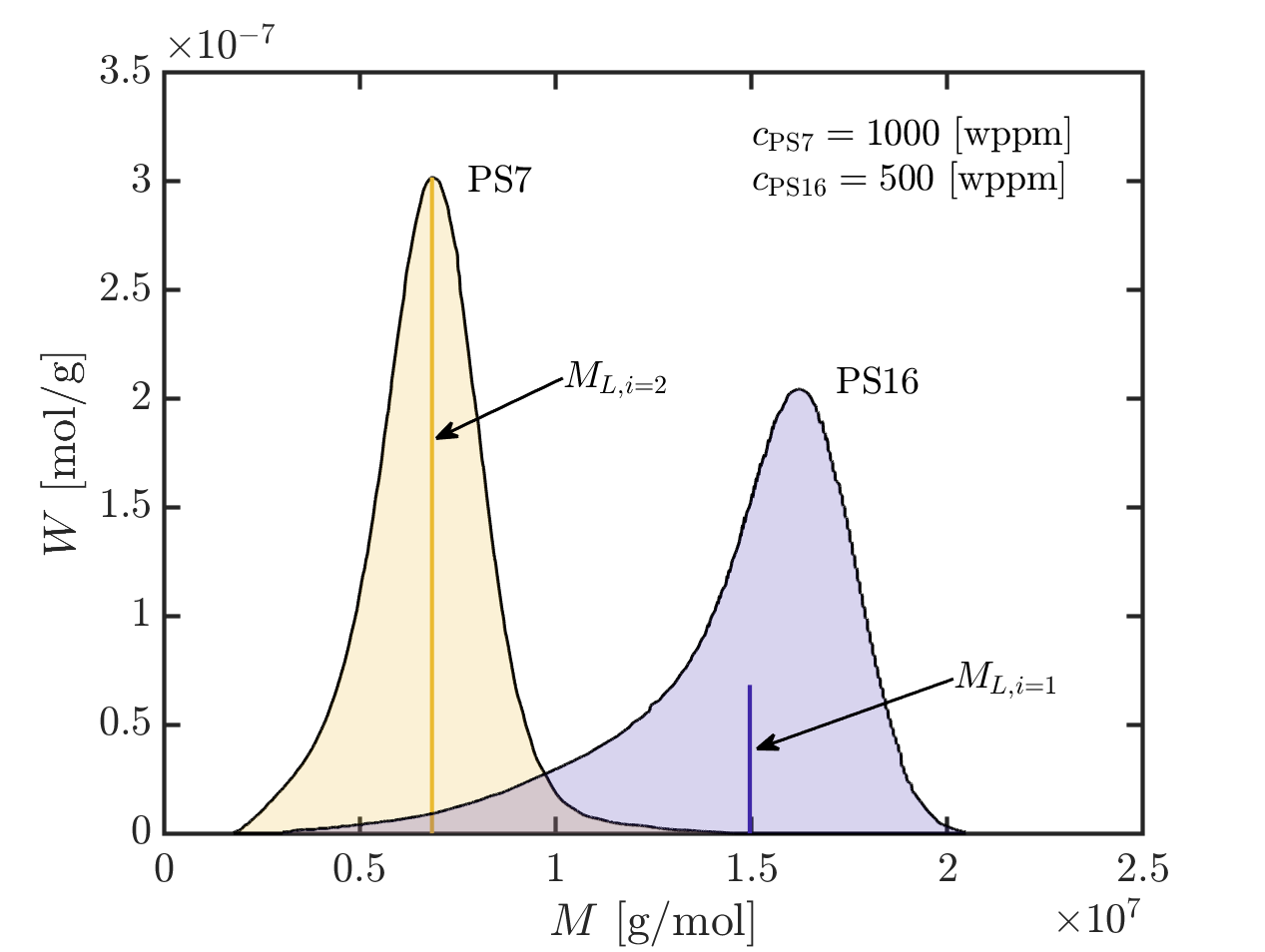}
    \caption{Measured molecular weight distributions for two narrowly-distributed polystyrene samples labeled PS7 and PS16  \cite{calabrese2025effects}. Each species has an extensibility-averaged molecular weight as defined by \eref{eq: continuous ext-weighted MW}, indicated by the location of the vertical lines. The line height is proportional to the relative number-density $n_i$ of each species (see \eref{eq: number density} and surrounding discussion) . Species are labeled in descending order of molecular weight by the index $i$.}
    \label{Fig1}
\end{figure}

This idea was recently explored systematically by Calabrese et al. \cite{calabrese2025effects}, who performed CaBER experiments on mixed solutions of two different polystyrene samples in dioctyl phthalate (DOP), each with distinct but overlapping molecular weight distributions as shown in \fref{Fig1} (referred to as PS7 and PS16 for the `low' and `high' molecular weight samples respectively).
From time-resolved measurement of the filament radius $a(t)$, they calculated the effective elongational strain rate as per \eref{eq: definition of tau_EC}, where $\dot{\varepsilon}_\mathrm{EC}$ was defined as the lowest measured value of the elongation rate throughout the experiment (this method is equivalent to a local exponential fit to the filament radius).
At sufficiently low concentrations of polymer, varying either the PS7 or PS16 concentrations was found to affect the measured $\tau_\mathrm{EC}$, but at higher concentrations of PS16, adding additional PS7 was shown to have virtually no impact on the experimentally-measured $\tau_\mathrm{EC}$.
They explained these results by arguing that only a portion of the molecular weight distribution of the polymers in solution is stretched prior to the onset of the EC regime.
During the initial IVC regime, the extension rate increases monotonically with time in their experiments.
Since, for a given $\dot{\varepsilon}$, the highest-molecular-weight chains with the longest $\tau_i$ experience the largest effective $W\!i_i \equiv \dot{\varepsilon} \tau_i$, these chains are the first to be elongated as the extension rate increases, and hence are the first to contribute to the total elastic polymer stress.
As the extension rate increases, lower-molecular-weight chains progressively begin contributing stress.
This continues until the tensile stress carried by the polymers is sufficient to induce the EC regime, and hence there is no `need' for the lower-molecular-weight chains to stretch.
Therefore, at sufficiently high concentrations of PS16, all the tensile stress in the thinning thread is carried by the PS16 polymer molecules to the exclusion of PS7, and hence the measured timescale is an inherent function of the total concentration and concentration distribution of the polymer sample in solution.

In this paper, we seek to explore this analysis computationally by explicitly incorporating polydispersity into the classic analysis of Entov and Hinch \cite{entov1997effect} for filament thinning of dilute solutions of polymer chains.
The chains are represented by FENE springs, with relaxation times set by the Zimm model, which incorporates preaveraged hydrodynamic interactions (HI).
Our model directly calculates the evolving polymeric stress contribution for each polymer species to the overall filament force balance, allowing systematic study of the time-evolution of the influence of each species.
We validate our model against the recent experimental results of Calabrese et al. \cite{calabrese2025effects} for mixtures of two different molecular weight samples of polystyrene, as described above and illustrated in \fref{Fig1}. 
These species are labelled PS7 and PS16 in line with the notation of Calabrese et al., corresponding to the modal molecular weight of the two distributions.
Our description of the filament thinning is necessarily approximate, as it does not incorporate a spatiotemporal fluid dynamical description using full 2D axisymmetric simulations (see for example Refs.~\citenum{zinelis2024fluid, ardekani2010dynamics}), nor do we seek to exactly reproduce the polymer conformation in extensional flow using simulations (see Refs.~\citenum{schroeder2004effect, prabhakar2017effect, prabhakar2016influence, hsieh2005prediction}). 
Instead, we primarily wish to illustrate, using a one dimensional time dependent simulation, the key features of polydispersity.
In particular, the relationship between the molecular weight distribution, the onset and evolution of stretch for different molecular weight species, and the resulting effects on the EC thinning timescale $\tau_\mathrm{EC}$ that is measured by experiments.

This paper is organised as follows.
We first present the model, starting with the filament force balance, followed by our procedure for determining the relevant model parameters by matching linear viscoelastic properties with experiments.
Next we detail the full nonlinear equation for the evolution of the tensile stress difference in the polydisperse polymer solution.
We then compare our model with experimental results for the PS7 and PS16 species individually, particularly contrasting the observable `elastocapillary timescale' $\tau_\mathrm{EC}$ with the actual polymer relaxation time $\tau$ in light of recent discussion in the literature \cite{clasen2006dilute, aisling2024importance, gaillard2024beware, hu2025revealing}.
As will be seen, the Zimm relaxation spectrum alone does not properly capture the experimental data except at very low values of $c/c^*$, and so we add in a conformation-dependent drag term to account for changes in the effective relaxation time of a polymer solution at finite concentration as polymer chains are stretched and begin to interact.
We then show that this model can reproduce the key experimental findings for polydisperse solutions, namely that at moderate concentrations of the high molecular-weight species, the influence of the low molecular-weight species on the thinning dynamics becomes negligible.
By examining the contribution to the elastic stress of each polymer species, we show that the low molecular weight species have a lower effective Weissenberg number on account of their lower relaxation time. 
If the Weissenberg number for a particular species falls below that required for the coil-stretch transition (i.e. $W\!i < 0.5$), the stress contribution from this species rapidly decays to zero.
In this way, we can define an analogue of \eref{eq: ext-weighted MW} which properly accounts for the set of all species which contribute to the tensile stress difference in the filament during the key elastocapillary regime.
Finally, we discuss the influence of experimental parameters such as plate radius ($a_0$) and initial sample pre-stretch (which is set by the aspect ratio of the CaBER instrument after plate separation) on the filament thinning predictions of our model.

\section{Methods}
\label{Methods}

Before we continue, we will first carefully define what we mean by `species' in this context. 
In a real-world setting, as for the experiments of Calabrese et al. \cite{calabrese2025effects}, one would purchase $N_i$ different polymer samples, each with a narrowly-dispersed and  distinct normalised mass-probability density function  $W_i(M)$ (not to be confused with the weight-fraction of each species $w_i$, which is just a number and not a function). 
In our case, this is the two samples referred to as PS7 ($i=2$) and PS16 ($i=1$) with molecular weight distributions $W_2(M)$ and $W_1(M)$ as shown in \fref{Fig1}.
Each of these $N_i$ polymer samples is added to the solvent at a mass-concentration $c_i$ (expressed throughout this paper in weight parts per million (wppm)), so that the weight-fraction of a particular sample is given by:
\begin{equation}
\label{eq: weight fraction}
    w_i = \frac{c_i}{\sum^{N_i}_i c_i}
\end{equation}
For our purposes in this paper (and as in \eref{eq: ext-weighted MW}), we idealise each of these $N_i$ samples as a monodisperse polymer `species' with a characteristic molecular weight $M_{L,i}$ given by the continuous mass-probability density function analogue of \eref{eq: ext-weighted MW} for the extensibility-weighted molecular weight:
\begin{equation}
\label{eq: continuous ext-weighted MW}
    M_{L,i} = \left( \frac{\int_0^\infty W_i(M) M^{1 + \nu} dM}{\int_0^\infty W_i(M) dM} \right)^\frac{1}{1 + \nu}
\end{equation}
where $M_{L,i}$ is the extensibility-weighted molecular weight of `species' $i$.
The number-density (units of polymer chains per volume) of species $i$ is given by:
\begin{equation}
\label{eq: number density}
    n_i = \frac{c_i \cdot \rho\cdot N_A}{10^6 M_{n,i}} 
\end{equation}
where $\rho$ is the fluid density, $N_A$ is Avogadro's constant, and $M_{n,i}$ is the number-averaged molecular weight of species $i$
\begin{equation}
    \frac{1}{M_{n,i}} = \int_0^\infty \frac{W_i(M)}{M}\mathrm{d}M.
\end{equation}
The overall extensibility-weighted molecular weight of the solution will then be given by \eref{eq: ext-weighted MW}, with $M_i \equiv M_{L,i}$ as from \eref{eq: continuous ext-weighted MW}, and $w_i$ as from \eref{eq: weight fraction}. 
Throughout the majority of this paper, this is what we mean by `species', namely the samples PS7 ($i=2$) and PS16 ($i=1$).

However, in reality, there is no true distinction between the $N_i$ different `species' (these are narrowly-distributed, but not monodisperse, polymer samples from a commercial supplier) --- they of course both contain the same linear polystyrene chains and are characterised by (slightly) overlapping molecular weight distributions. 
In this sense, each possible distinct polymer chain length in the final mixed disperse solution would be its own `species', each characterised by a particular mass-fraction. 
This is clearly unwieldy to manage from an analytical point of view, and so throughout most of this paper we set $N_i = 2$, with each `species' corresponding to the two samples used in the experiments of Calabrese et al. with known $W_i(M)$ and $c_i$.
In \sref{sec: stress evolution} for Figs. \ref{Fig7} and \ref{Fig8}, to more precisely analyse the elastic stress evolution in the filament, we instead combine the two mass-probability density function distributions into a single bidisperse distribution and then break this broad distribution into $N_j$ equally-spaced bins, each of which now represents a particular `sub-species' (with the index $j$ used instead of $i$ when referring to this new method of splitting into sub-species).
This second method of defining species is detailed in \aref{Appendix_splitting_MW}.
Using this latter, more complete approach means that our model can be applied to describe filament thinning flows of any polydisperse dilute polymer solution, as long as the underlying molecular weight distribution is known/measured (and the total concentration of polymer in solution is dilute, i.e. non-interacting). 

\subsection{Stress balance for thinning filament}

We wish to describe the thinning of a cylindrical filament with radius $a(t)$ under the action of capillary, viscous, inertial, and polymer stresses.
As shown by other authors \cite{mckinley2005visco, entov1997effect, zinelis2024fluid}, one can derive from these terms a one-dimensional version of a force balance for the extension rate within the filament:
\begin{equation}
\label{eq: force balance}
    0 = \frac{C}{4} \dot{\varepsilon}^2 a^2 \rho + \Delta \sigma_p + 3 \eta_s \dot{\varepsilon} - (2 X - 1) \frac{\chi}{a}
\end{equation}
where $C$ is a numerical constant quantifying the strength of inertial effects, $\Delta \sigma_p (t)$ is the polymer contribution to the tensile stress difference (for convenience we use the notation $\Delta$ because this is a stress difference between axial and radial stresses $\Delta \sigma (t) = \sigma_{zz} - \sigma_{rr}$ in cylindrical coordinates), $\eta_s$ is the solvent viscosity, $X$ is a numerical factor characterising the slenderness of the filament shape (conservation of mass for a cylindrical fluid element being elongated in a shear-free flow gives $X=1$, which is henceforth assumed throughout), and $\chi$ is the surface tension of the fluid.
The first term represents the inertial contribution to the stress, the second term the polymer contribution, the third term the viscous contribution from the solvent, and the fourth the influence of surface tension. 

The thinning of the filament due to surface tension generates a uniaxial extensional flow, and an initially cylindrical material element undergoing such a flow then evolves as:
\begin{equation}
\label{eq: filament radius evolution}
    \dot{a} = -\frac{1}{2} \dot{\varepsilon} a
\end{equation}
Equations (\ref{eq: force balance}) and (\ref{eq: filament radius evolution}) can be integrated to obtain the evolving filament radius as a function of time, and they represent a `0 + 1 dimensional' equation in space and time. 
The kinematics of this description are accurate once a stable cylindrical filament is formed, but necessarily is only an approximation of the full fluid evolution at early times, when the role of the end plates may be important, depending on the aspect ratio (see \aref{Appendix_pre_stretch} as well as Spiegelberg et al. \cite{spiegelberg1996role}).

This set of equations is typically non-dimensionalised by setting the characteristic length scale to the initial filament radius, $a_0$, and the characteristic time scale to the `Rayleigh time' $t_R = a_0/\sqrt{\chi/\rho a_0}$.
If we denote the dimensionless radius by $a^*$ and dimensionless tensile stress difference by ${\Delta \sigma_p}^*$, we obtain:
\begin{equation}
    0 = C \dot{a^*}^2 - (2X-1)\frac{1}{a^*} - \frac{6 \eta_s}{\sqrt{\rho a_0 \chi}} \frac{\dot{a^*}}{a^*} + \frac{\eta_p a_0}{\tau \chi} {\Delta \sigma_p}^*
\end{equation}
Here we have assumed that the polymer stress can be characterised by a polymeric viscosity $\eta_p$ (i.e. the polymer contribution to the viscosity in the limit of vanishingly small extension rates), and a relaxation time $\tau$, which can also be expressed as an elastic modulus $g = \eta_p/\tau$.
As we will see, the polymer solutions we investigate here have a well-defined and experimentally observable viscosity $\eta_p$, but they do not have a single relaxation time $\tau$.
Rather, they are characterised by a spectrum of relaxation times, and so the use of a single $\tau$ should be viewed as a single mode approximation of the full dynamics \cite{entov1997effect}.
If we further define a total viscosity $\eta = \eta_s + \eta_p$, we have the following equation:
\begin{equation}
    0 = C \cdot \dot{a^*}^2 - (2X-1)\frac{1}{a^*} - 6 Oh \cdot \beta \frac{\dot{a^*}}{a^*} + \frac{1}{Ec} \Delta \sigma_p^*
\end{equation}
with the following dimensionless parameters: Ohnesorge number $Oh$ (the relative contributions of viscous and inertial forces) and elastocapillary number $Ec$ (the ratio of capillary pressure to polymer stress),
\begin{subequations}
\begin{equation}
    Oh = \frac{\eta}{\sqrt{\rho \chi a_0}}
\end{equation}
\begin{equation}
    Ec = \frac{\tau \chi}{\eta_p a_0} \equiv \frac{\chi}{g a_0}
\end{equation}
plus a viscosity ratio
\begin{equation}
    \beta = \frac{\eta_s}{\eta_p + \eta_s} \equiv \frac{\eta_s}{\eta}.
\end{equation}
\end{subequations}
Appropriate values of $Oh$, $Ec$ and $\beta$ for single-component solutions of 1000ppm PS7 and 200ppm PS16 are given in \tref{tab:parameters}.

It remains to describe the evolution of the polymer stresses within the thinning filament.
We begin by characterising the linear viscoelastic properties of the polystyrene solutions employed in the experiments.
We will then describe a full nonlinear evolution equation for the polymer stress, further incorporating the finite extensibility of the polymer chains, and conformation-induced changes to the effective chain-solvent friction that arise as the molecules begin to stretch.

\subsection{Linear viscoelasticity}

\begin{figure*}[ht]
    \centerline{
    \begin{tabular}{c c}
        \includegraphics[width=7.5cm,height=!]{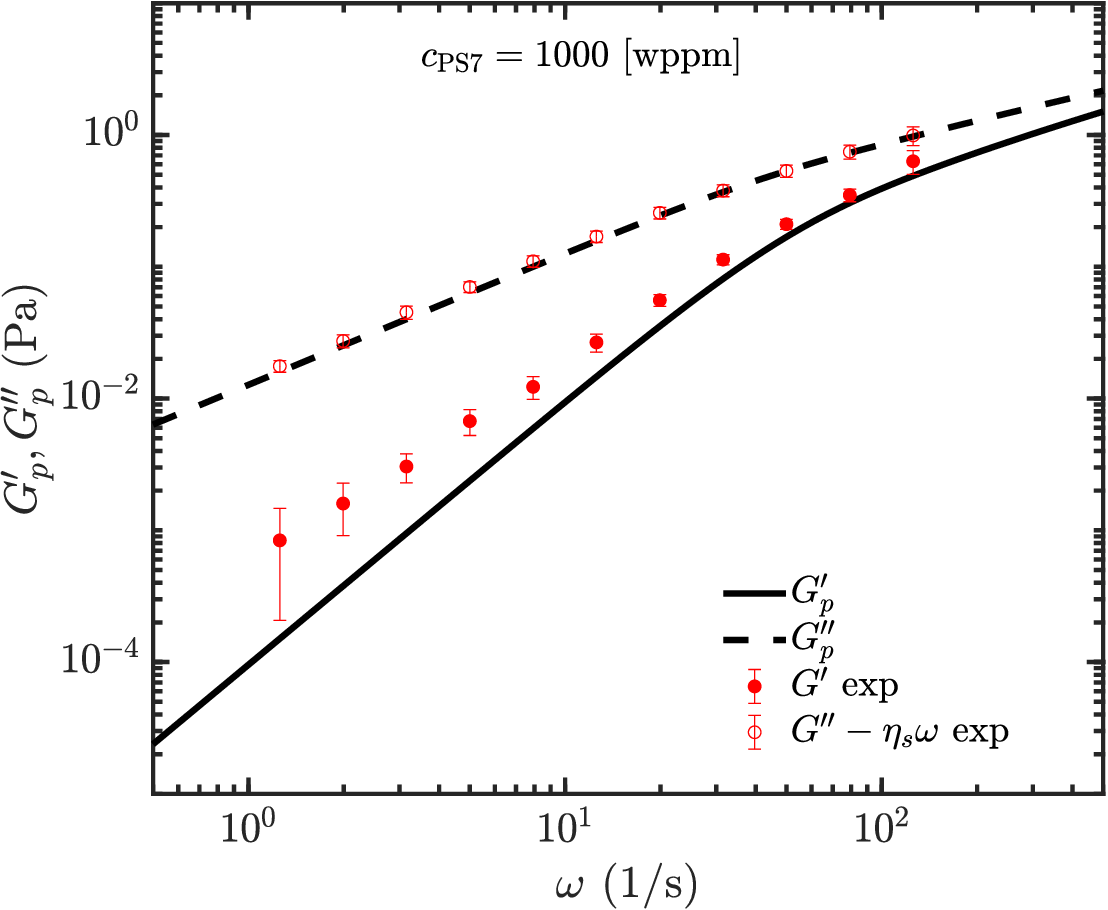} & \includegraphics[width=7.5cm,height=!]{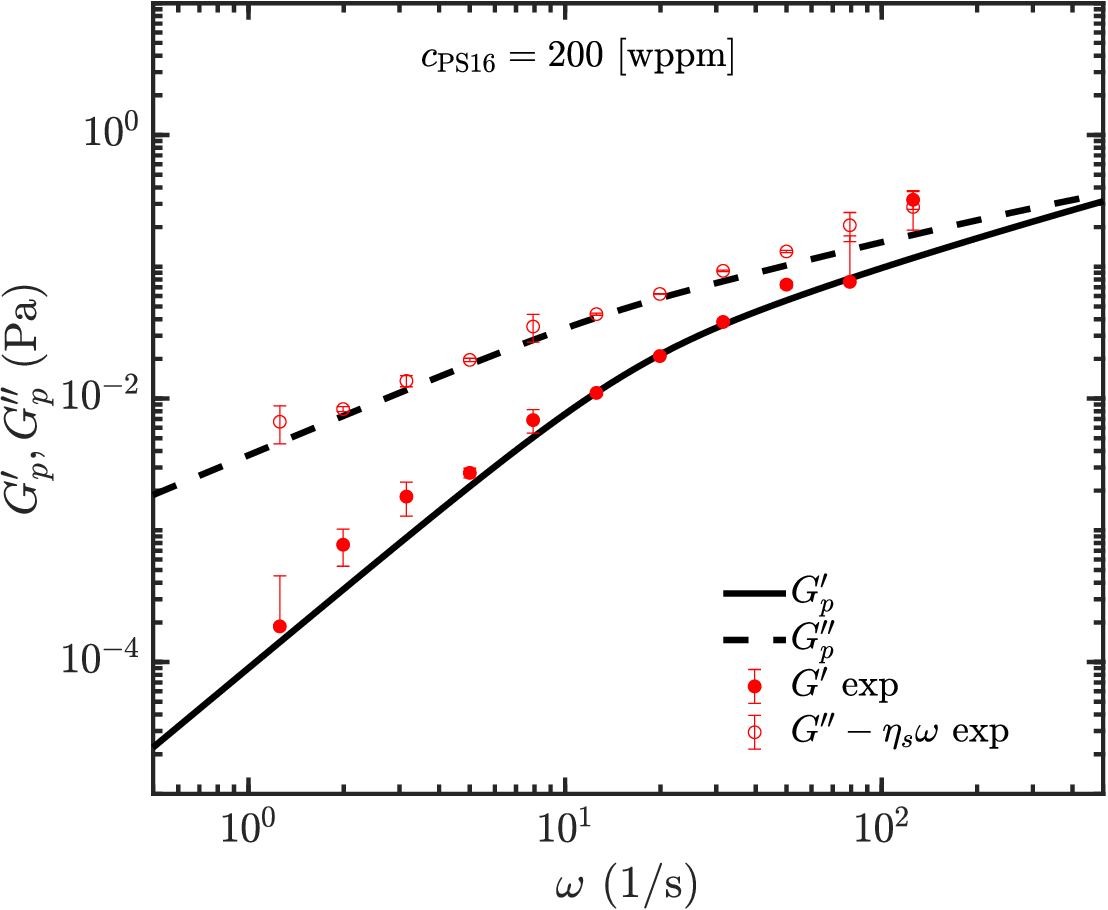} \\
        (a) & (b) \\
    \end{tabular}
    }
    \caption{The polymer contribution to the storage ($G_p'$) and loss ($G_p''$) moduli for a dilute solution of polystyrene in DOP. The solvent contribution $\eta_s \omega$ to the loss modulus has been subtracted away from the experimental data. Lines are computed using \eref{eq: G' G''} with parameters as given in the text and \tref{tab:parameters}. Symbols are experimental data, with the standard deviation over triplicate measurements given as error bars. (a) PS7 solution (species $i=2$) at 1000 ppm by weight, such that $c/c^*_{i=2} = 0.26$. (b) PS16 solution (species $i=1$) at 200 ppm by weight, such that $c/c^*_{i=1} = 0.08$.}
    \label{Fig2}
\end{figure*}

The linear viscoelastic response of dilute solutions of polystyrene in a $\theta$-solvent, such as the dioctyl phthalate (DOP) used in Ref.~\citenum{calabrese2025effects}, is known to be well-described in the linear regime by the Zimm model \cite{rubinstein2003polymer, johnson1970infinite, zimm1956dynamics}.
Each long flexible polymer molecule is conceptualized as a chain of $N_s$ independent freely jointed springs which connect together $N_s + 1$ massless point particles \cite{zimm1956dynamics, bird1987dynamics}.
The point particles represent effective centers of hydrodynamic resistance and move under the actions of Brownian motion and external forces resulting from fluid flow, while the springs represent the entropic elasticity of the polymer molecule as it is deformed and generate a restoring force.
One considers an ensemble of such chains, for which a Fokker-Planck equation can be derived for the probability density of the chain conformation \cite{bird1987dynamics}.
This ensemble must be non-interacting, or dilute, which is quantified through the polymer overlap concentration $c_i^*$ given by \cite{rubinstein2003polymer}:
\begin{equation}
\label{eq: overlap concentration}
    c_i^* = \frac{M_i}{N_A \left(\frac{2 \sqrt{\left\langle R^2 \right\rangle_{i, \mathrm{eq}}}}{\sqrt{6}} \right)^3} \cdot \frac{10^6}{\rho}
\end{equation}
where $\sqrt{\langle R^2 \rangle_{i, \mathrm{eq}}}$ is the root mean square end to end distance of the chain at equilibrium (defined for the current sample later in \eref{eq: RMS end to end}), and the $10^6/\rho$ term is added to maintain the current convention of expressing concentrations in wppm.
Dilute solutions are those with $c_i \ll c_i^*$.
This Fokker-Planck system of equations is approximately solved through pre-averaging of the hydrodynamic interactions between the point particles and decomposition into normal modes \cite{zimm1956dynamics, Ottinger1996}.
One obtains $N_s$ evolution equations for the normal modes of the total stress arising from the chain species $i$, each of which is characterised by a timescale $\tau_{is}$.
These timescales are the relaxation times or Zimm relaxation time spectrum of the chain.
It can then be shown that the contribution of a particular polymer species $i$ with a Zimm relaxation time spectrum $\tau_{is}$ to the (overall polymer contribution to the) storage and loss moduli $G'_p$ and $G''_p$ from all the normal spring modes ($N_s$) is \cite{rubinstein2003polymer}:
\begin{subequations}
\label{eq: G' G''}
\begin{equation}
    G'_{p,i} = g_i \sum^{N_s}_s \frac{\omega^2 \tau_{is}^2}{1 + \omega^2 \tau_{is}^2}
\end{equation}
\begin{equation}
    G''_{p,i} = g_i \sum^{N_s}_s \frac{\omega \tau_{is}}{1 + \omega^2 \tau_{is}^2}
\end{equation}
\end{subequations}
\begin{equation}
\label{eq: polymer modulus}
    g_i = n_i k_\mathrm{B} T
\end{equation}
where $\omega$ is the oscillatory frequency and $g_i$ is the polymer modulus which is proportional to the number density of polymers $n_i$ of species $i$.
The total storage and loss moduli for the dilute solution of bidisperse polymer chains is then:
\begin{subequations}
\begin{equation}
    G' = \sum_i^{N_i} G'_{p, i}
\end{equation}
\begin{equation}
    G'' = \eta_s \omega + \sum_i^{N_i} G''_{p, i}
\end{equation}
\end{subequations}
where for mixtures of different species $i$ at low values of $c_i/c_i^*$, the moduli $G'_{p,i}$ and $G''_{p,i}$ add linearly, and we also include the solvent contribution $\eta_s \omega$.

In the Zimm model, the longest relaxation time ($s = 1$ by convention, also known as the Zimm time $\tau_Z$) can be found from intrinsic viscosity measurements as \cite{rubinstein2003polymer}:
\begin{equation}
    \tau_{Z, i} \equiv \tau_{i,s=1} = \frac{1}{U_{\tau \eta}}\frac{M_{i} [\eta] \eta_s }{N_A k_\mathrm{B} T}
\end{equation}
where $k_\mathrm{B}$ is Boltzmann's constant, $T = 298$ K is the temperature, $U_{\tau \eta}$ is a universal constant equal to $\approx 2.36$ in a $\theta$-solvent \cite{Larson2005review}, while the intrinsic viscosity $[\eta]$ is given by the Mark-Houwink relation:
\begin{equation}
    [\eta]_i = K M_i^\alpha
\end{equation}
where $\alpha = 3 \nu - 1$, $\nu = 0.5$ for PS in DOP (which acts as a $\theta$-solvent at the experimental temperature $T = 25^\circ C$), and $K = 0.08$ cm$^3$/g, all taken from literature \cite{PolymerHandbook1999}.
The relaxation times for modes $s\ge2$ can be approximated via Thurston's relation for the equilibrium pre-averaged contributions of the normal modes of the Zimm bead-spring chain \cite{bird1987dynamics}, 
\begin{equation}
    \tau_{is} = \frac{\tau_{i,s=1}}{s^{2+p}}
\end{equation}
where $p$ is the exponent in Thurston's formula, given by:
\begin{equation}
    p = -1.40 {h^*}^{0.78}
\end{equation}
Here $h^*$, the strength of hydrodynamic interactions, is essentially a free parameter, which we set to $h^* = 0.26$ in order to obtain a good fit to the linear viscoelastic data, in line with literature values \cite{Larson2005review, johnson1970infinite}.
We show these comparisons in \fref{Fig2}, where we have measured the linear viscoelasticity of the PS7 and PS16 samples independently.
As described earlier, the approximate molecular weight $M_i$ of each species is given by the extensibility-weighted average $M_{L,i}$ as in \eref{eq: continuous ext-weighted MW}.

\begin{table}[htbp]
\centering
\caption{Model parameters and their values used in the simulations, separated into continuum level fluid and geometric parameters (above) and polymer chain parameters (below). Note that for polystyrene, $C_\infty$ is defined per carbon-carbon bond, not per styrene monomer \cite{rubinstein2003polymer}.}
\label{tab:parameters}
\begin{tabularx}{\columnwidth}{@{}lclX@{}}
\toprule
\textbf{Parameter} & \textbf{Value} & \textbf{Units} & \textbf{Description} \\
\midrule
$C$ & $144$ & --- & Inertial constant \\
$\rho$ & $985$ & kg\,m$^{-3}$ & Fluid density \\
$\eta_s$ & $0.059$ & Pa\,s & Solvent viscosity \\
$X$ & $1$ & --- & Filament shape factor \\
$\chi$ & $0.031$ & N\,m$^{-1}$ & Fluid surface tension \\
$a_0$ & $0.62$ & mm & Initial filament radius \\
$Oh_\mathrm{PS7}$ & $0.52$ & --- & Ohnesorge no. for 1000ppm PS7 \\
$Oh_\mathrm{PS16}$ & $0.46$ & --- & Ohnesorge no. for 200ppm PS16 \\
$Ec_\mathrm{PS7}$ & $130$ & --- & Elastocapillary no. for 1000ppm PS7 \\
$Ec_\mathrm{PS16}$ & $1400$ & --- & Elastocapillary no. for 200ppm PS16 \\
$\beta_\mathrm{PS7}$ & $0.82$ & --- & Viscosity ratio for 1000ppm PS7 \\
$\beta_\mathrm{PS16}$ & $0.94$ & --- & Viscosity ratio for 200ppm PS16 \\
\midrule
$U_{\tau \eta}$ & $2.36$ & --- & Universal polymer constant \\
$T$ & $298$ & K & Temperature \\
$K$ & $0.08$ & cm$^3$\,g$^{-1}$ & Mark--Houwink constant \\
$\alpha$ & $3\nu - 1$ & --- & Mark--Houwink exponent \\
$\nu$ & $0.5$ & --- & Excluded volume exponent \\
$h^*$ & $0.26$ & --- & Chain HI parameter \\
$h^*_K$ & $0.027$ & --- & Segmental HI parameter \\
$l$ & $0.154$ & nm & Carbon--carbon bond length \\
$M_\mathrm{o}$ & $104$ & g\,mol$^{-1}$ & Styrene monomer molecular weight \\
$C_{\infty}$ & $9.7$ & --- & Characteristic ratio \\
$c^*_\mathrm{PS7}$ & $3920$ & wppm & Overlap concentration for PS7 \\
$c^*_\mathrm{PS16}$ & $2590$ & wppm & Overlap concentration for PS16 \\
$\tau_{2,1} | \tau_{Z, \mathrm{PS7}}$ & $0.015$ & s & Longest relaxation time for PS7 \\
$\tau_{1,1} | \tau_{Z, \mathrm{PS16}}$ & $0.047$ & s & Longest relaxation time for PS16 \\
\bottomrule
\end{tabularx}
\end{table}

\subsection{FENE-PM model}

We now extend this description into a differential equation for the nonlinear evolution of the polymer stress tensor, which additionally accounts for the finite extensibility of the chains.
The conformational ensemble for each species is evolved according to the FENE-PM model (where the PM implies both a Peterlin closure approximation and a mean averaging over the extensibility of each spring), originally introduced by Wedgewood et al. \cite{wedgewood1991finitely} and given in its present form by Anna et al. \cite{anna2001interlaboratory}:
\begin{equation}
\label{eq: FENE-PM}
    \dot{\bm{A}}_{is} = \bm{\kappa \cdot A_{is}} + \bm{A_{is} \cdot \kappa^T} + \frac{f_i}{\tau_{is}} \left(\bm{\delta} - \bm{A}_{is} \right)
\end{equation}
\begin{equation}
\label{eq: FENE extensibility factor}
    f_i = \frac{L_i^2}{L_i^2 + 3 N_s - \sum_{s=1}^{N_s} \mathrm{Tr} \bm{A}_{is}}
\end{equation}
where $\bm{A}_{is}$ is the normalised second moment of chain extension for species $i$ and mode $s$, $\bm{\kappa}$ the homogeneous velocity gradient tensor $\nabla \bm{v}$, $\tau_{is}$ the relaxation time of mode $s$ of species $i$, $\bm{\delta}$ the identity tensor, $f_i$ the nonlinear spring factor of species $i$, and $L_i$ the finite extensibility of species $i$ (an expression for which will be given shortly).
The second index $s$ refers to the normal mode decomposition from $s = 1$ for the longest Zimm relaxation time of chain $i$ to $s = N_s$ for the shortest relaxation time.
Note that we place the nonlinear spring factor $f_i$ on both the spring and Brownian terms, rather than on the spring force alone as in Refs. \citenum{wedgewood1991finitely, anna2001interlaboratory}; the two forms coincide in the linear viscoelastic limit and in the terminal regime, and we have verified that they give indistinguishable results for the cases considered here.

This model assumes that the chain is made up of $N_s$ finitely extensible springs, where the spring nonlinearity at large extensions is given as an average over the ensemble of chains and springs (the sum over $\mathrm{Tr} \bm{A}_{is}$ in the expression for the nonlinear force $f_i$).
This averaging separates the FENE-PM chain from the FENE-P chain model, where each spring has its own nonlinearity $f_{is}$ corresponding to $\bm{A}_{is}$ \cite{wedgewood1991finitely}. 
Both models employ the Peterlin closure approximation to obtain a final ODE for the evolution of the ensemble \cite{bird1987dynamics}.
In the full FENE model without the Peterlin closure, each \textit{chain} in the ensemble has a separate $f_{is}$, such that the $f \bm{A}$ term essentially remains as $\left\langle f \bm{QQ} \right\rangle$ (where $\bm{Q}$ is the end-to-end vector of an individual spring), and so a closed-form ODE cannot be derived and one must numerically solve the full Fokker-Planck equation or the equivalent stochastic differential equation for the chain conformation \cite{Ottinger1996}.
The FENE-PM approximation further allows one to separate the evolution equation into a series of normal modes, greatly simplifying computations compared to the FENE-P chain \cite{wedgewood1991finitely}.

Because of this normal mode decomposition, $\bm{A}_{is}$ does not refer to the extension of a particular spring, but rather the averaged conformation associated to a particular normal mode with relaxation time $\tau_{is}$ (details can be found in Ref.~\citenum{wedgewood1991finitely}, where our $\bm{A}_{is}$ for a particular species corresponds to $\bm{\sigma}_j$ in their notation, and we use a slightly different construction for our $f_i$, or $Z$ in their notation). 
\eref{eq: FENE-PM} describes the evolution of the normal modes of the chain, such that $\bm{A}_{is}$ is the conformation of the $s$th mode of a chain with a molecular weight of $M_i$ (which in the present analysis of extension-dominated flow is the extensibility-averaged molecular weight of the chain, $M_{L,i}$).
Since the nonlinear spring factor $f_i$ is a sum over the $N_s$ modes, it is possible for any single mode $s$ to be fully `extended' ($\mathrm{Tr} \bm{A}_{is} \rightarrow L_i^2 + 3N_s$) while the others are completely relaxed --- only the total extension of the chain is constrained, not that of each individual spring.
In the linear viscoelastic limit (where $f_i \rightarrow 1$ and $\bm{A}$ is infinitesimally perturbed from isotropy),  \eref{eq: G' G''} is recovered.
When $N_s = 1$, the equations reduce to the regular FENE-P dumbbell model.

As before, $i$ refers to the species (in descending order of molecular weight), while 
\begin{equation}
\label{eq: L calc}
    L_i^2 = 3\frac{R_{i, \mathrm{max}}^2}{\langle R^2 \rangle_{i, \mathrm{eq}}} - 1
\end{equation}
For polystyrene, both $R_{\mathrm{max},i}$ and $\langle R^2 \rangle_{i, \mathrm{eq}}$ have been quantified in $\theta$-solvents, with:
\begin{subequations}
\begin{equation}
    R_{i, \mathrm{max}}^2 = \frac{4 l^2 M_i^2}{M^2_\mathrm{o}}
\end{equation}
\begin{equation}
\label{eq: RMS end to end}
    \langle R^2 \rangle_{i, \mathrm{eq}} = \frac{2 C_\infty l^2 M_i}{M_\mathrm{o} }
\end{equation}
\end{subequations}
where $l = 0.154$nm is the length of a C-C bond, $M_\mathrm{o} = 104$ g/mol is the molecular weight of the styrene monomer, and $C_\infty$ is the characteristic ratio \cite{PolymerHandbook1999}. 
Note that since $R_{i,\mathrm{max}}^2$ is proportional to the number of Kuhn steps squared ($N_{i,K}^2$) while $\langle R^2 \rangle_{i, \mathrm{eq}}$ is proportional to $N_{i,K}$, the number of Kuhn steps in the chain is $(L_i^2 + 1)/3$.

We require only one final relation, which is that for the total polymer stress, given as:
\begin{equation}
\label{eq: polymer stress}
    \bm{\sigma}_p = \sum_i g_i f_i \sum_s \bm{A}_{is}
\end{equation}
where $g_i$ is the elastic modulus of species $i$ as in \eref{eq: G' G''}.
When Eqs. \ref{eq: FENE-PM} and \ref{eq: polymer stress} are specialised to a shear-free time-dependent uniaxial extension, we have:
\begin{subequations}
\label{eq: conformation evolution elongation}
\begin{equation}
    \dot{A}_{is,zz} = 2 \dot{\varepsilon} A_{is,zz} - \frac{f_i}{\tau_{is}} \left(A_{is,zz} - 1 \right)
\end{equation}
\begin{equation}
    \dot{A}_{is,rr} = - \dot{\varepsilon} A_{is,rr} - \frac{f_i}{\tau_{is}} \left(A_{is,rr} - 1 \right)
\end{equation}
\end{subequations}
\begin{equation}
\label{eq: polymer stress difference}
    \Delta \sigma_p = \sum_i g_i f_i \sum_s (A_{is,zz} - A_{is,rr})
\end{equation}
where $A_{is,rr}$ and $A_{is,zz}$ are the radial ($rr$) and axial ($zz$) components of the polymer conformation tensor for mode $s$ of the species $i$. 

\subsection{Conformation-dependent drag}
\label{Appendix_CDD}

When a polymer chain stretches in flow, it no longer interacts hydrodynamically with the solvent as a coil, but instead as an elongated ellipsoidal filament, with a corresponding change in friction \cite{Larson2005review, prabhakar2016influence, prabhakar2017effect}.
To incorporate this effect into our simulations, we use a simplified version of the approach of Prabhakar and coworkers \cite{prabhakar2017effect}.
Using polymer blob theory \cite{Pincus1976} and hydrodynamic calculations, they were able to develop an expression for the evolution in the overall chain drag coefficient as a function of chain size, concentration, extension, and local segmental friction for a single-mode, monodisperse suspension of dumbbells.
Adapting this model to a multi-mode, polydisperse solution is not trivial and beyond the scope of the current work.
Therefore, we have incorporated only the expressions for the effective extension-dependent friction of isolated chains, and applied it only to the evolution equation for $\bm{A}_{i1}$ (the longest mode for each species). 
We focus on the hydrodynamic effects on $\bm{A}_{i1}$ as this relaxation mode encodes the largest-scale changes in the conformation of species $i$ (analogous to a low-order spherical harmonic expansion) and will have the largest influence upon the stress at long times.

We will not re-derive the model, only giving the final result.
The reader is referred to the work of Prabhakar and coworkers for details \cite{prabhakar2017effect}.
The modified longest relaxation time is given by:
\begin{equation}
\label{eq: CDD tau}
    \hat{\tau}_{i1}(\bm{A}_\mathrm{i1}) = \left[\frac{\alpha_H(h_K^*, (L_i^2 + 1)/\mathrm{Tr}\bm{A}_\mathrm{i1})}{\alpha_H(h_K^*, (L_i^2 + 1)/3)}\right] \left(\frac{\sqrt{\mathrm{Tr}\bm{A}_\mathrm{i1}/3}}{\ln\left(2 \mathrm{Tr}\bm{A}_\mathrm{i1}/3\right)}\right)
\end{equation}
where $\alpha_H$ is the hydrodynamic shielding effectiveness ratio, given by:
\begin{equation}
    \alpha_H(h^*_K, N) = \frac{6 \pi^{3/2} h^*_K}{P_1(N) + 2 \sqrt{\pi} h^*_K P_2(N)}
\end{equation}
which involves the fitted quadratic functions $P_1$ and $P_2$:
\begin{subequations}
\begin{equation}
    P_1 = 0.91 - \frac{0.02}{\sqrt{N}} + \frac{0.10}{N}
\end{equation}
\begin{equation}
    P_2 = 1.6 - \frac{1.1}{\sqrt{N}} - \frac{0.5}{N}
\end{equation}
\end{subequations}
and $(L_i^2 + 1)/3$ (denoted $N_K$ in Prabhakar et al. \cite{prabhakar2016influence}) is the number of Kuhn steps in the chain, $\mathrm{Tr}\bm{A}_\mathrm{i1}/3$ ($N_t$ in Prabhakar et al. \cite{prabhakar2016influence}) is the number of `tension blobs' in the chain, and $h^*_K$ is the segmental hydrodynamic interaction parameter, which is not necessarily equal to $h^*$ for the overall chain and so is an independent parameter. 
When fully extended, $N_K \equiv N_t$, so a `tension blob' is the same size as a Kuhn step, while at equilibrium $N_t = 1$.
The final evolution equation for the first (longest) mode of species $i$ with the same $N_s$ and concentration $n_i$ as the equivalent species without conformation-dependent drag is then:
\begin{equation}
    \dot{\bm{A}}_{i1} = \bm{\kappa \cdot A_{i1}} + \bm{A_{i1} \cdot \kappa^T} + \frac{f_i}{\hat{\tau}_{i1}(\bm{A}_\mathrm{i1})} \left(\bm{\delta} - \bm{A}_{i1} \right)
\end{equation}
while the relaxation times of the other modes (describing the shorter relaxing modes of the chain) have the original, constant $\tau_{is}$.

\section{Results and Discussion}
\label{Results}

\begin{figure}[t]
    \centerline{
    \begin{tabular}{c}
        \includegraphics[width=7.5cm,height=!]{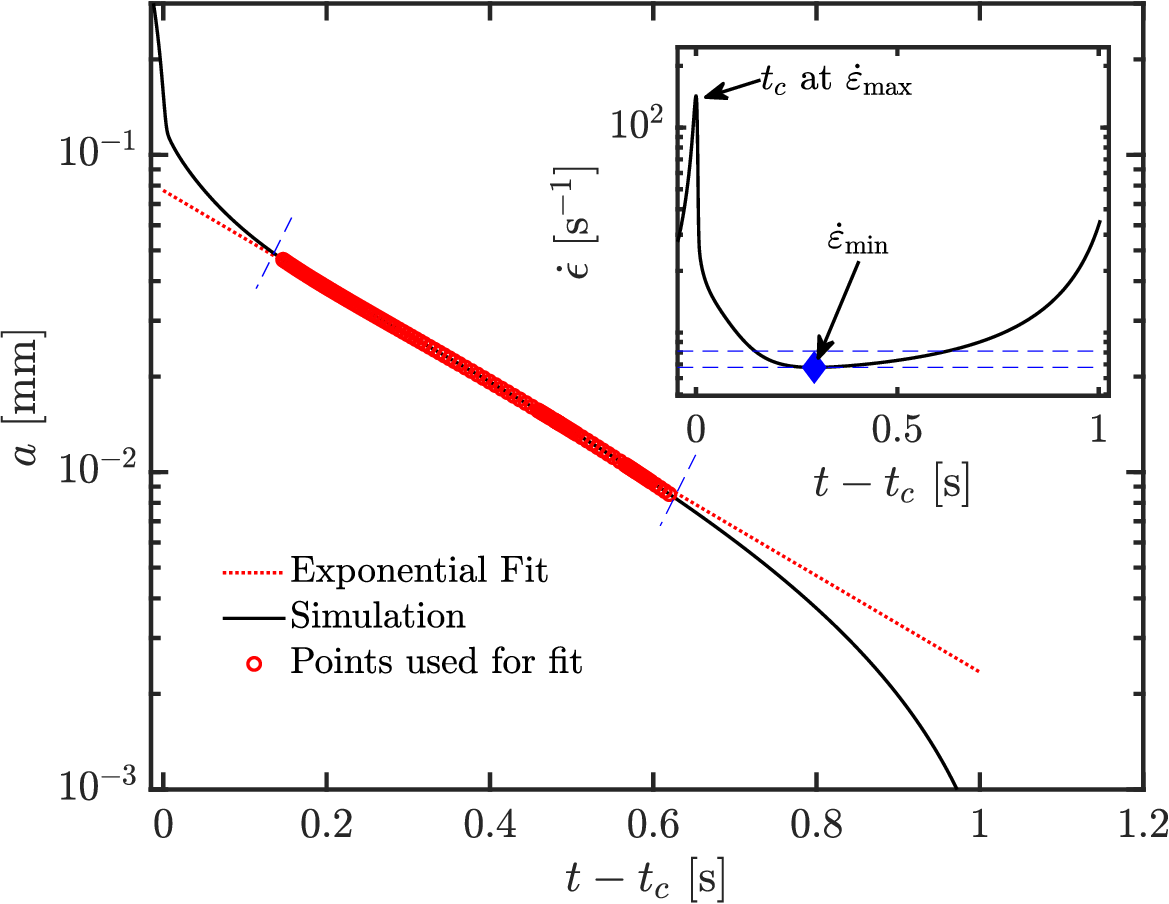} 
    \end{tabular}
    }
    \caption{Method for calculating elastocapillary timescale $\tau_\mathrm{EC}$. The main plot shows a simulation of the filament radius over time (black line), while the inset shows the corresponding extension rate during the same simulation. In the inset, the range of extension rates within 10\% of the local minimum $\dot{\varepsilon}_\mathrm{min}$ are shown by blue dashed lines, with the minimum extension rate as a blue diamond. The corresponding times are labeled as blue dashed lines in the main figure, and points between these limits are used for a local exponential fit as per \eref{eq: fitting exponential} to find $\tau_\mathrm{EC}$. Times are shifted by $t_c$, the time corresponding to the peak in the extension rate at the end of the IVC regime.}
    \label{Fig3}
\end{figure}

To compute the evolution of the filament over time, we solve the system of ordinary differential equations \eqref{eq: force balance}, \eqref{eq: filament radius evolution}, \eqref{eq: conformation evolution elongation}, and \eqref{eq: polymer stress difference} simultaneously using Matlab's ode15s routine. 
Since we don't compute the full fluid dynamical solution for the initial necking in the liquid filament, we regard the inertial constant $C$ and the initial filament radius $a_0$ as fitting parameters, which we set to $C = 144$ and $a_0 = 0.62$ mm throughout.
These values are physically reasonable - the minimum radius for the onset of necking in the liquid filament in the experiments was $\approx 0.6$ mm, and full 2D axisymmetric simulations of filament formation and breakup using a FENE-P model found values for the constant $C$ in the range of 30 to 150 (see Ref.\citenum{zinelis2024fluid}). 
All model parameters for polystyrene in DOP solution are as per \sref{Methods} \cite{PolymerHandbook1999}.
Parameter values in our simulations are summarised in \tref{tab:parameters}, as well as the Ohnesorge number ($Oh \sim 1$, which indicates that the initial thinning takes place in a regime where inertial and viscous forces are roughly balanced \cite{zinelis2024fluid, mckinley2005visco}), and the overlap concentration \eref{eq: overlap concentration} for each species.
We extract the effective filament thinning timescale $\tau_\mathrm{EC}$ through an exponential fit to the evolution of the filament radius in the EC regime, as shown in \fref{Fig3}.
Specifically, we fit:
\begin{equation}
\label{eq: fitting exponential}
    a = a_1 \exp\left( -\frac{t - t_c}{3 \tau_\mathrm{EC}} \right)
\end{equation}
where $a_1$ and $\tau_\mathrm{EC}$ are the fitting constants, and $t_c$ is the time corresponding to the local maximum of the elongation rate (which occurs only after significant elastic stresses have developed and marks the transition between the initial IVC regime and the subsequent EC regime). 
The inset to \fref{Fig3} displays the time evolution of the elongational strain rate as per \eref{eq: filament radius evolution}.
One can also identify the EC timescale as $\tau_\mathrm{EC} = 2/(3\dot{\varepsilon}_\mathrm{min})$ as highlighted in the inset.
This was the method used in Calabrese et al.\cite{calabrese2025effects}.

\subsection{Thinning of PS7 and PS16 individually}

\begin{figure}[ht]
    \centerline{
    \begin{tabular}{c}
        \includegraphics[width=7.5cm,height=!]{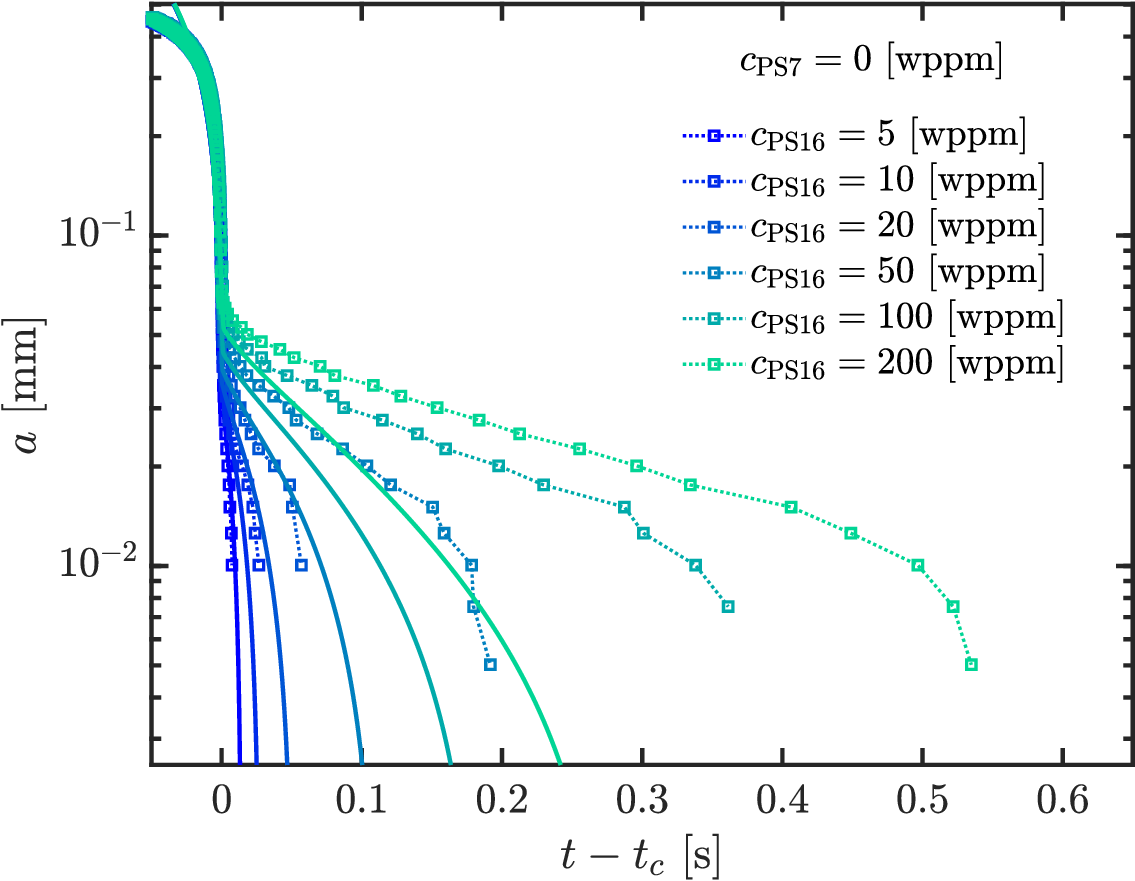} \\
        (a) \\
        \includegraphics[width=7.5cm,height=!]{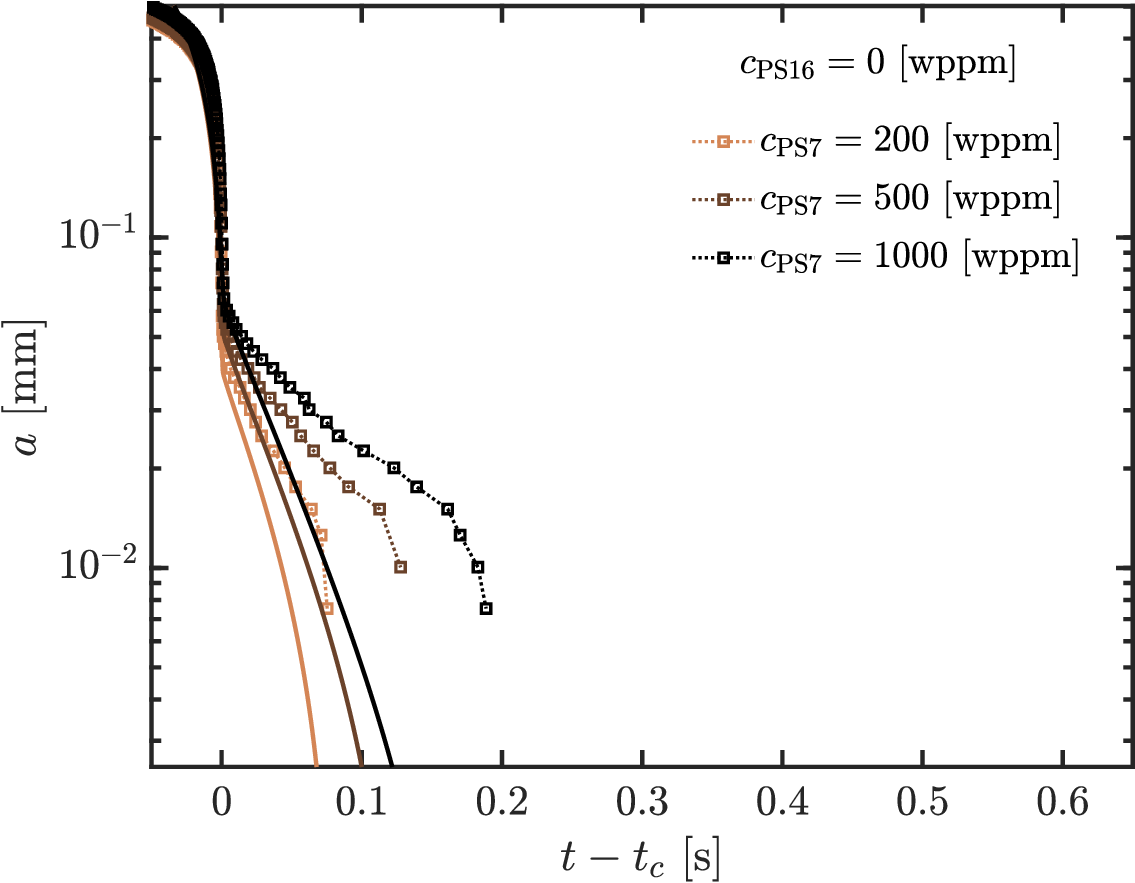} \\
        (b) \\
    \end{tabular}
    }
    \caption{Evolution of the mid-filament radius $a(t)$ during a CaBER experiment. $t_c$ is defined as the time at which the local maximum extension rate is achieved during the initial IVC thinning, as in \fref{Fig3}. Experimental results (symbols and dotted lines) are those of Calabrese et al. \cite{calabrese2025effects} for polystyrene in DOP solvent. Solid lines are predictions of our model without any conformation-dependent drag, as described in the text.}
    \label{Fig4}
\end{figure}

\begin{figure}[t]
    \centerline{
    \begin{tabular}{c}
        \includegraphics[width=8.5cm,height=!]{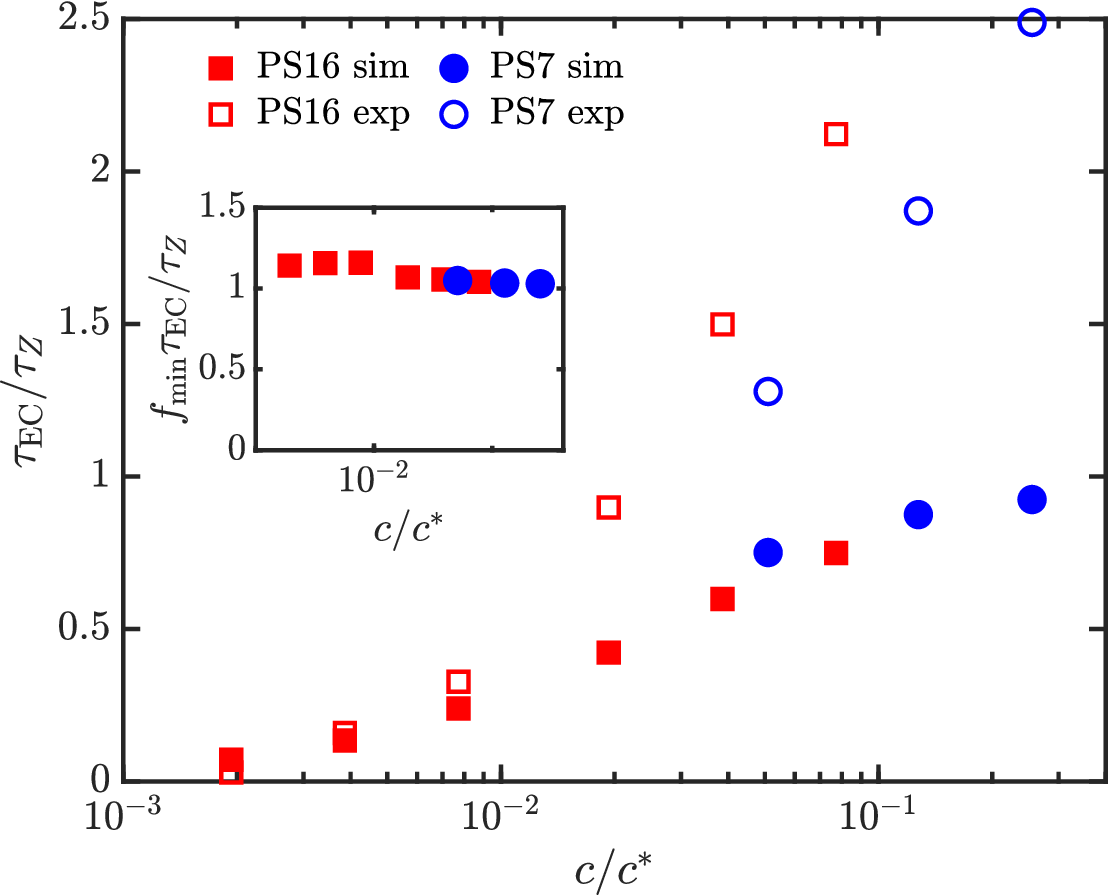} 
    \end{tabular}
    }
    \caption{Ratio of $\tau_\mathrm{EC}$ to $\tau_Z$ plotted against the dimensionless concentration $c/c^*$ for experiments (hollow points) and simulations (filled points). $\tau_\mathrm{EC}$ is calculated from a local exponential fit to the capillary thinning dynamics as described in \fref{Fig3}. The Zimm time $\tau_Z \equiv \tau_{i1}$ is the longest relaxation time for PS7 and PS16 respectively. In the inset, the same data is replotted using the effective timescale $\tau_Z/f_\mathrm{min}$, with $f_\mathrm{min}$ the value of $f$ calculated using \eref{eq: FENE extensibility factor} at the time corresponding to the minimum strain rate.}
    \label{Fig5}
\end{figure}

We first wish to examine the predictions of our model for the two polystyrene species in isolation, before considering mixtures.
We compare our numerical predictions with the experimental data of Calabrese et al. \cite{calabrese2025effects}, for both PS7 and PS16.
Our simulations do not yet incorporate conformation-dependent drag effects.
These comparisons are plotted in \fref{Fig4} for several concentrations of PS7 and PS16.
Our results roughly agree with experiment at low concentrations, but disagree at higher concentrations.
This effect becomes clearer if we plot the ratio of $\tau_\mathrm{EC}$ over the Zimm time $\tau_Z$ (obtained independently from linear viscoelastic measurements) against the effective concentration $c/c^*$, as in \fref{Fig5}.
Both the simulation and experiment show that at low concentrations, $\tau_\mathrm{EC}$ is considerably smaller than $\tau_Z$.
At higher concentrations, $\tau_\mathrm{EC}$ approaches $\tau_Z$ for the simulations, but increases above $\tau_Z$ in the experiments.

The cause of this behaviour in simulations is straightforward, namely that the equivalence between $\tau_\mathrm{EC}$ and $\tau_Z$ for the Oldroyd-B model only holds when there is no nonlinear spring factor, i.e. when $f = 1$ in FENE-P-type models ($L \rightarrow \infty$).
For $f > 1$, it is immediately clear from \eref{eq: conformation evolution elongation} that the `effective' relaxation time for a particular mode is $\tau_{is} / f_i$, as $\tau_{is}$ does not appear elsewhere, by itself, in the evolution equations.
This nonlinear correction is shown explicitly in the inset to \fref{Fig5}, where $\tau_Z$ is divided by the nonlinear spring factor $f_\mathrm{min}$ at the time corresponding to $\dot{\varepsilon}_\mathrm{min}$. 
In this case, one obtains a nearly constant ratio of $\tau_Z/f_\mathrm{min}$ to $\tau_\mathrm{EC}$, implying that most of the deviation observed computationally arises due to nonlinearity in the elastic spring force in the elongating polymer.
At lower concentrations, the elastic modulus of the chains in the solution is lower, and so the chains must stretch more to accumulate sufficient stress to `overcome' the viscous forces and balance capillarity in the EC regime.
Therefore, a solution at lower concentration will necessarily have a smaller effective relaxation time $\tau_Z/f_\mathrm{min}$ at the onset of the elastocapillary regime, and the value of $\tau_\mathrm{EC}$ determined from experimental observation of capillary thinning will be lower.
From these arguments, it is evident that $\tau_\mathrm{EC}$ can show a concentration dependence even without explicit chain-chain interactions, purely through the changes in the effective elastic modulus $g$ of the dilute system.

The extensibility dependence of $\tau_\mathrm{EC}$ for the classic Entov-Hinch model \cite{entov1997effect} has clearly been understood for some time\cite{clasen2006dilute, campo2010slow, wagner2015analytic}, but apparently only made explicit in several recent papers \cite{aisling2024importance, hu2025revealing, gaillard2024beware, gaillard2025does} and appears to not be widely appreciated.
For example, Clasen et al. \cite{clasen2006dilute} showed in 2006 that by fitting $\tau_Z$ to experimental data using the Entov-Hinch model, one obtains a constant fitted $\tau_Z$ at low concentrations, but not a constant $\tau_\mathrm{EC}$, as is consistent with the results above.
A later 2010 paper by Campo-Deaño and Clasen \cite{campo2010slow} defined a characteristic concentration, $c_\mathrm{low}$, below which a distinct EC regime cannot be identified due to the chain nonlinearity.
Others have derived semi-analytical expressions for the capillary thinning dynamics which clearly show the dependence of the measured exponential thinning rate on both the elastic modulus and the extensibility of the polymer chains in solution \cite{wagner2015analytic, hu2025revealing, gaillard2025does}.
This is also consistent with experimental studies showing that $\tau_\mathrm{EC}$ is influenced by both concentration and extensibility \cite{calabrese2024polymers, dinic2020flexibility, tirtaatmadja2006drop}.
We note also that since it is the value of the finite extensibility at the onset of the EC regime that sets the measured elastocapillary time constant $\tau_\mathrm{EC}$ in our model, it can also be influenced by experimental parameters such as the initial pre-stretch during filament formation (as in Ref.~\citenum{aisling2024importance}), or the initial filament radius (as in Ref.~\citenum{gaillard2024beware}). 
We discuss these effects in more detail in \sref{sec: pre stretch and plate diameter}.

At higher (but still small relative to $c^*$) concentrations, it is also clear from \fref{Fig5} that there is a systematic deviation of the dilute bead-spring simulations from experiments (even though $c/c^* < 1$), and an increase of $\tau_\mathrm{EC}$ beyond $\tau_Z$.
The same effect is seen in small amplitude oscillatory shear (SAOS) measurements \cite{clasen2006dilute}, where the measured value of the relaxation time begins to increase as neighboring chains begin to interact.
However, chain-chain interactions may appear at a lower $c/c^*$ in strong extensional flows than for measurements performed in the linear viscoelastic regime due to the so-called self-concentration effect, in which the effective overlap concentration of a stretched chain is smaller than for a coil \cite{prabhakar2016influence, stoltz2006concentration, dinic2019macromolecular, dinic2020flexibility}.
There are also suggestions that this dependence of viscoelastic properties on $c/c^*$ arises from polydispersity of the polymer sample \cite{calabrese2025effects}.

The physics of dynamic self-concentration remains an ongoing area of research which we will not attempt to account for here by including explicit concentration dependence into the individual spring relaxation times $\tau_{is}$ for each species.
Instead, we adjust the model to match experimental results at higher concentrations by incorporating the effects of conformation-dependent drag, where the effective relaxation timescale of the longest mode $\hat{\tau}_{i1}$ is a function of the chain extension $\mathrm{Tr}(\bm{A}_{i1})$ (see \eref{eq: CDD tau}).
The expressions used in this paper for $\hat{\tau}_{i1}$ are given in \sref{Appendix_CDD} and the origin of these additional physical effects is discussed in detail in Prabhakar et al. \cite{prabhakar2016influence}.

\subsection{Multiple species}

To compare with experimental results for mixtures of PS7 and PS16, we incorporate conformation-dependent drag as described in \sref{Appendix_CDD}.
Results are shown in \fref{Fig6}~(a) for a mixture of PS7 and PS16, where we have selected $h^*_K = 0.027$ to give excellent agreement for $c_\mathrm{PS16} = 200$ ppm.
The experimental results are again those of Calabrese et al. \cite{calabrese2025effects}, utilising the CaBER slow retraction method.
The scaling of the elastocapillary timescale $\tau_\mathrm{EC}$ for our model as a function of the concentrations of PS7 and PS16 is given in \fref{Fig6}~(b), alongside the equivalent experimental results.
The model reproduces the key qualitative trend, namely that small amounts of PS16 can have an outsized influence upon the observed thinning compared to a lower-molecular weight species.
In fact, in both experiments and simulations, at a concentration of only $c_\mathrm{PS16} \sim 50$ ppm there is very little difference in the observed thinning between $c_\mathrm{PS7} = 1000$ ppm and $c_\mathrm{PS7} = 0$ ppm, while for $c_\mathrm{PS16} \lesssim 50$ ppm, the concentration of $c_\mathrm{PS7}$ does have a significant effect.
This is in line with previous experimental findings \cite{calabrese2025effects, plog2005influence, palangetic2014dispersity}.
The implication is that the presence of PS16 affects the contribution of PS7 to the elastic stress, an idea which we will explore explicitly below.
However, our model fails to fully capture the dependence of $\tau_\mathrm{EC}$ on $c_\mathrm{PS7}$, likely due to the lack of explicit concentration dependence in our model (we see from \fref{Fig5} that the dimensionless concentration $c/c^*$ is higher for PS7 than for PS16).

\begin{figure}[ht]
    \centerline{
    \begin{tabular}{c}
        \includegraphics[width=7.5cm,height=!]{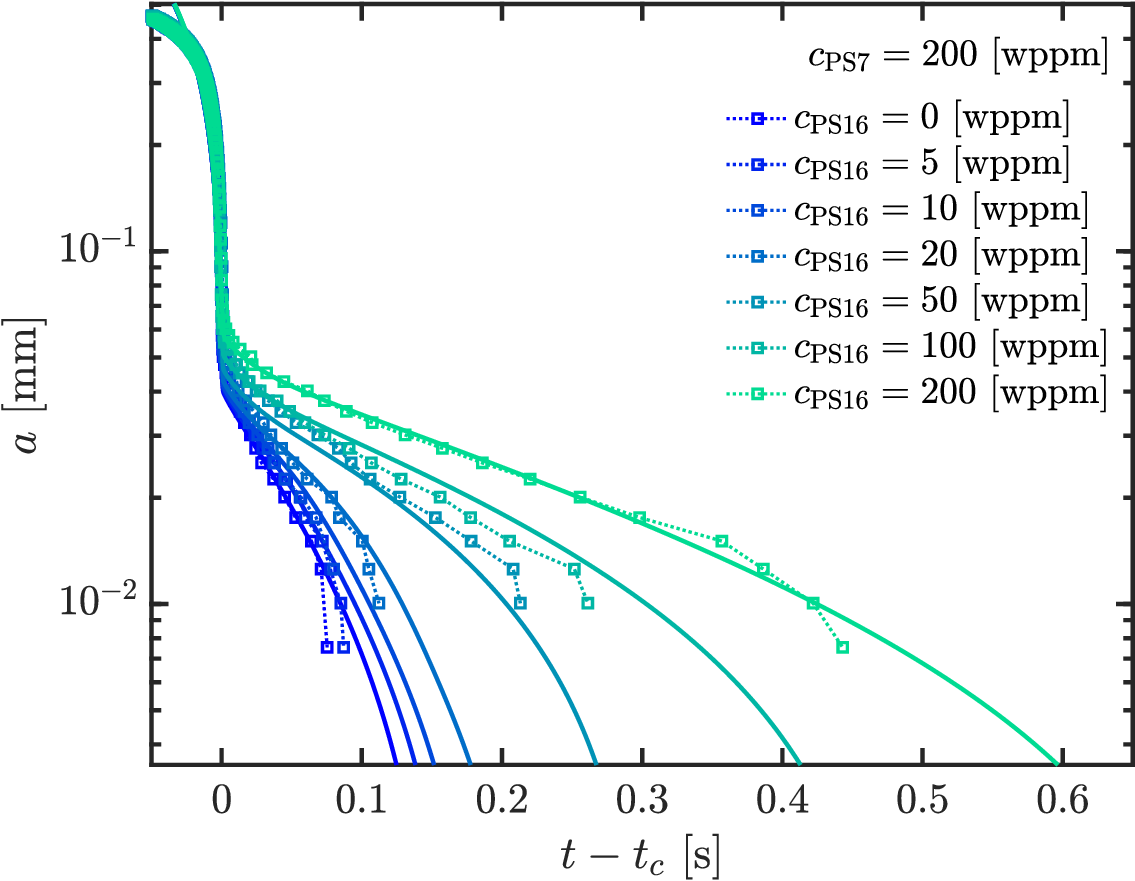} \\ 
        (a) \\
        \includegraphics[width=7.5cm,height=!]{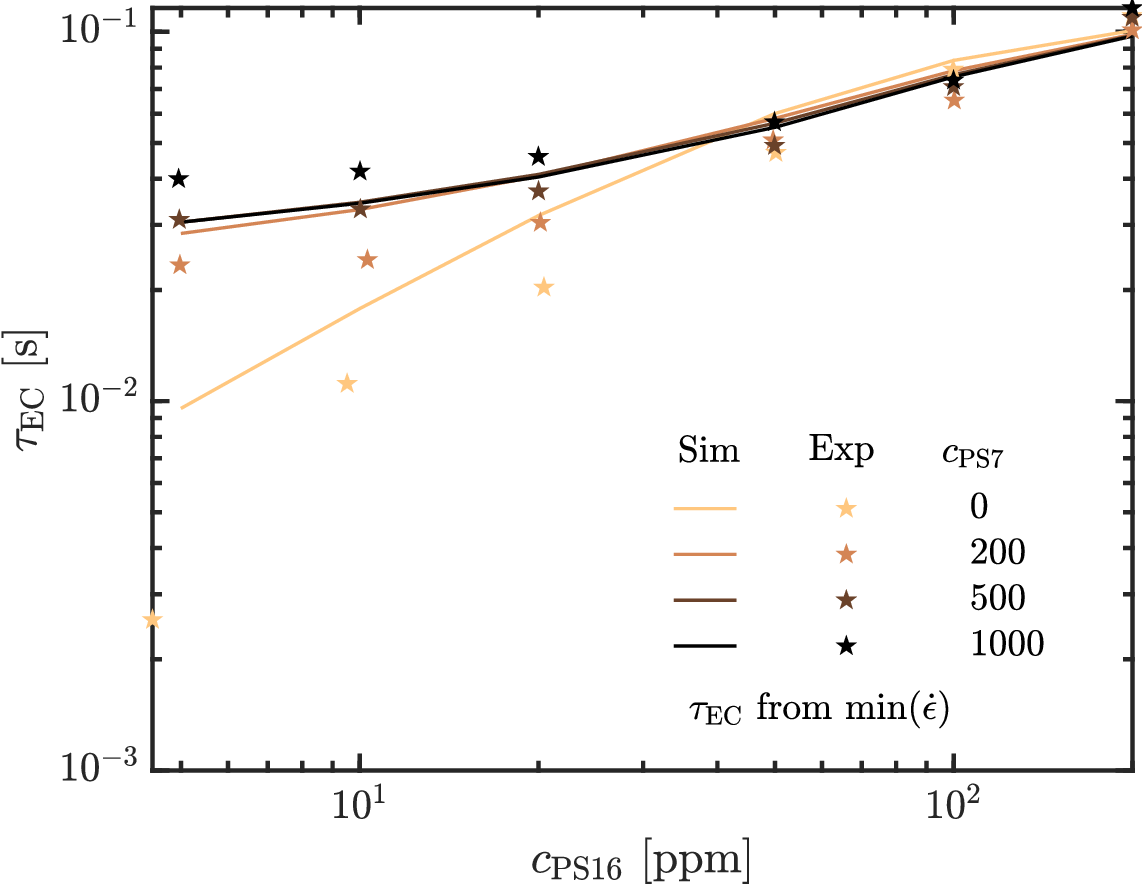} \\
        (b) 
    \end{tabular}
    }
    \caption{(a) Evolution of the mid-filament radius $a(t)$ during a CaBER experiment. $t_c$ is defined as the maximum in the extension rate during the initial IVC thinning. Experimental results (symbols and dotted lines) are those of Calabrese et al. \cite{calabrese2025effects} for polystyrene in DOP solvent, with mixtures of PS7 and PS16. Solid lines are predictions of our model with conformation-dependent drag included, as described in the text. (b) variation of $\tau_\mathrm{EC}$ for several concentrations of PS7 and PS16, calculated as described in \fref{Fig3}. Stars are experimental results, while lines are results of our simulations.}
    \label{Fig6}
\end{figure}

To rationalize the outsized influence of high-MW species, we can return to the analysis of Entov and Hinch \cite{entov1997effect}.
As we have seen, the relaxation time of a particular species scales as $\tau \sim M^{1.5}$, so longer chains have higher relaxation times.
Prior to the terminal linear regime when finite extensibility effects play a role, the extension rate in the filament slowly decreases as it asymptotically approaches a constant value (cf. inset to \fref{Fig3}), so that some portion of the modes fall below the critical coil-stretch transition (with Weissenberg number $W\!i = \dot{\varepsilon} \tau_{is} < 0.5$), and their contribution to the total polymer stress in the filament decreases over time rather than rising.
Eventually, only some small portion of the molecular weight distribution is contributing to the elastic stress that balances capillarity.
In the following section, we use our model to interrogate this effect in detail by directly calculating the time-dependent stress contribution from each molecular weight fraction.

\subsection{Stress evolution}
\label{sec: stress evolution}

\begin{figure*}[ht]
    \centerline{
    \begin{tabular}{c c}
        \includegraphics[width=8.5cm,height=!]{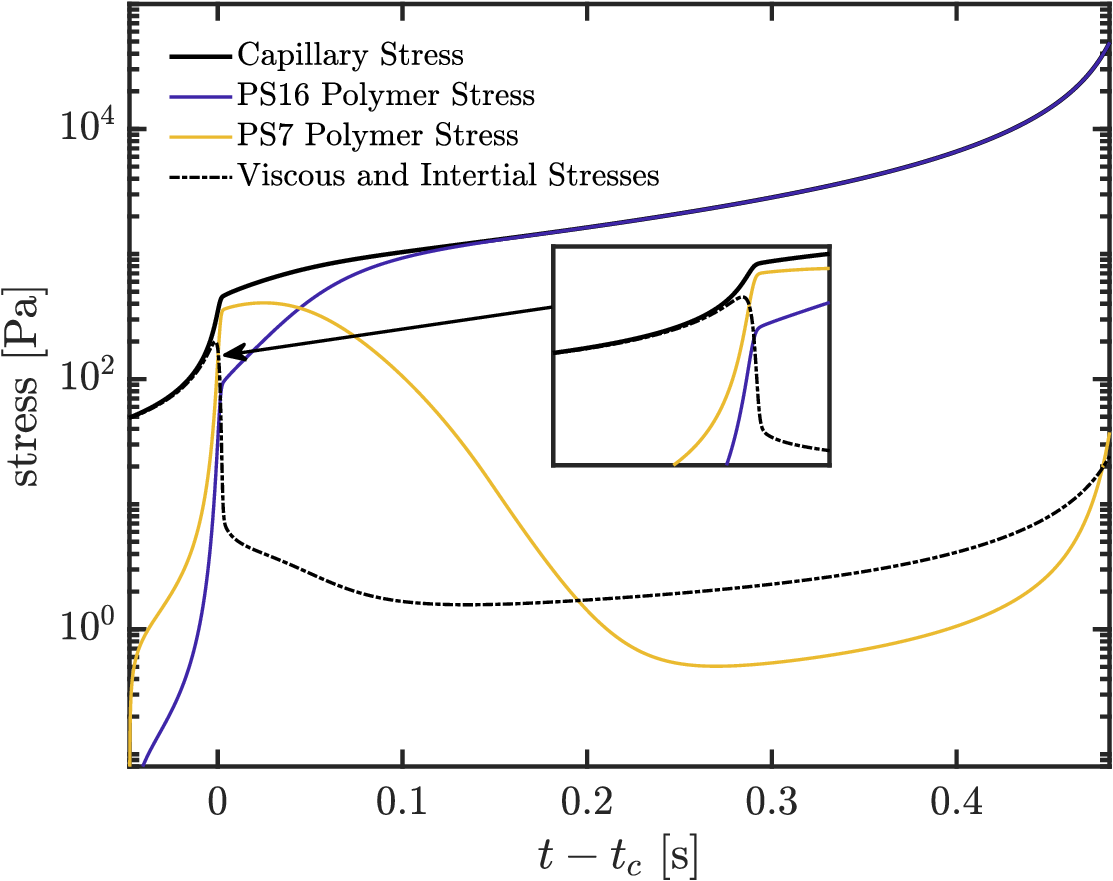} & \includegraphics[width=8.5cm,height=!]{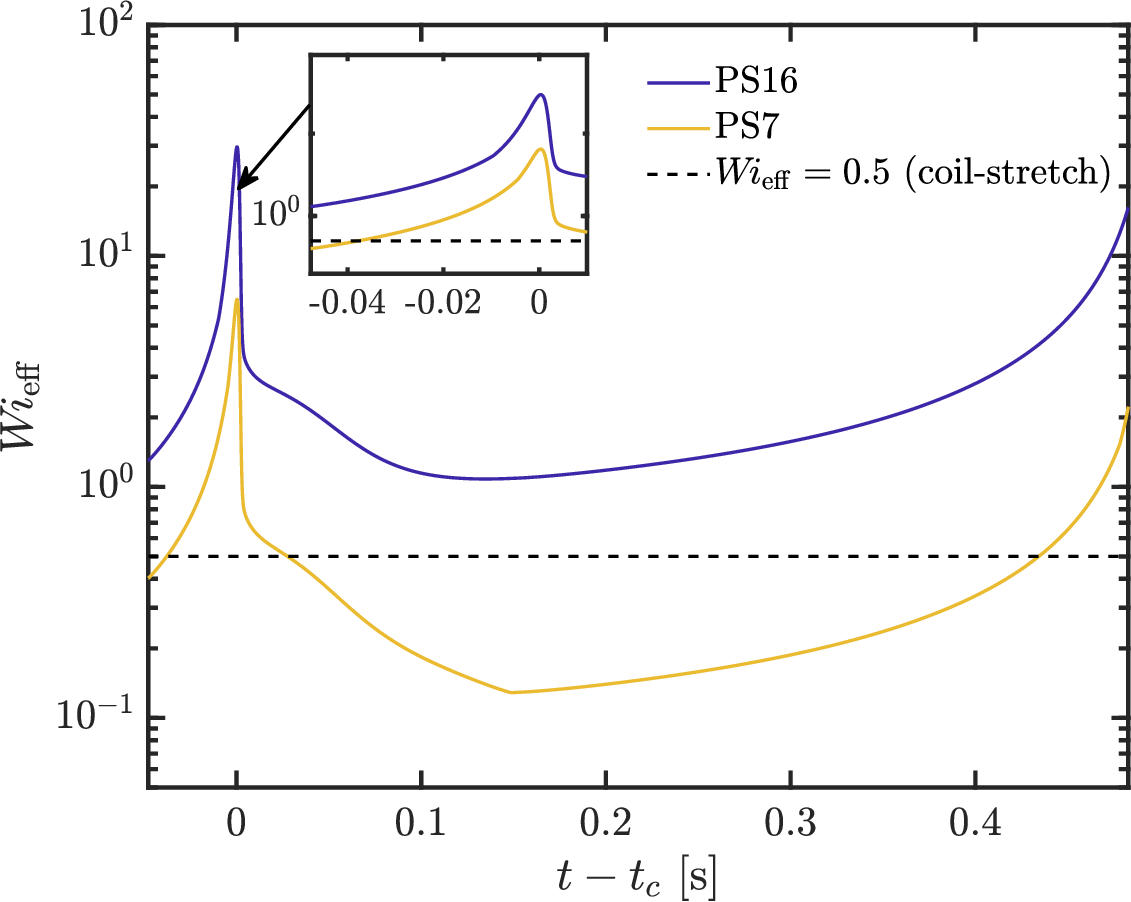} \\
        (a) & (b) \\
        \includegraphics[width=8.5cm,height=!]{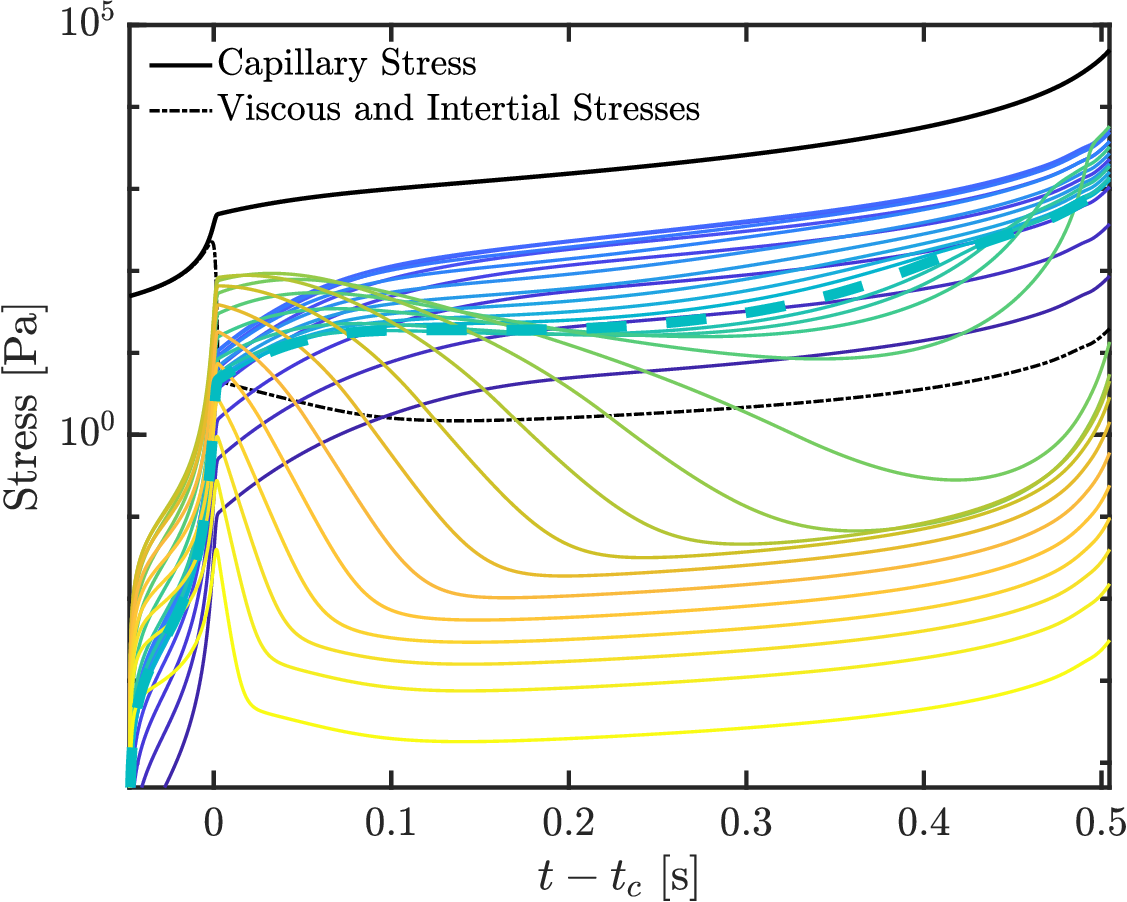} & \includegraphics[width=8.5cm,height=!]{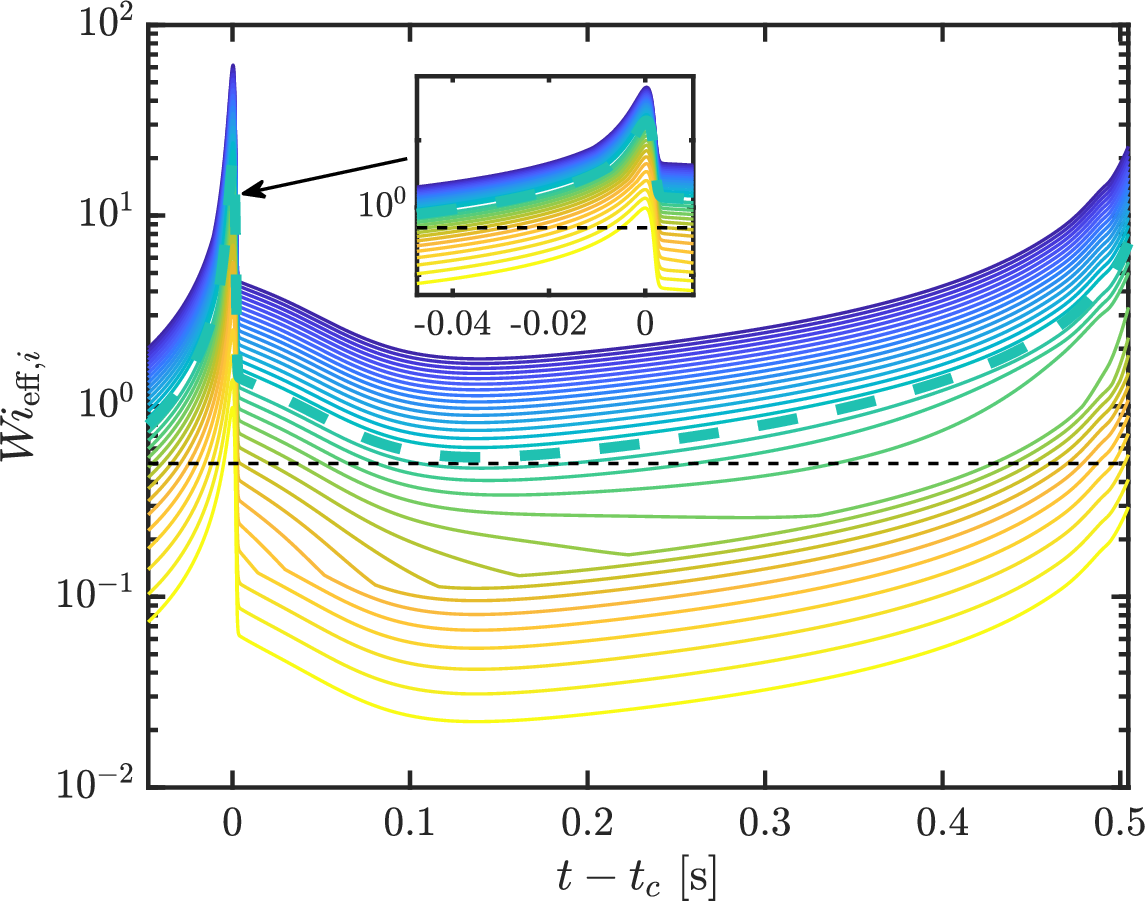} \\
        (c) & (d) \\
    \end{tabular}
    }
    \caption{Contributions to stress and effective extension rate of each polymer species during filament thinning for a mixture of 1000ppm PS7 and 500ppm PS16. Weissenberg number $W\!i_{\mathrm{eff},i} = \dot{\varepsilon} \hat{\tau}_{i1}$ for each molecular weight. Dashed line gives the coil-stretch transition at $W\!i = 0.5$. Inset shows behaviour at early times. (a) \& (b) Blue corresponds to PS16 and yellow to PS7 as per \fref{Fig1}.  (c) \& (d) Stress evolution when the molecular weight distribution is split into $N_j = 30$ sub-species, as detailed in \aref{Appendix_splitting_MW}. Colours represent the molecular weight of each sub-species (identically to \fref{FigA1}), with blue for the highest molecular weight ($M_{1}= 20 \mathrm{MDa}$) to yellow for the lowest molecular weight ($M_{30}= 0.22 \mathrm{MDa}$). The thicker dashed line is that sub-species which just touches the coil-stretch transition at $W\!i = 0.5$.}
    \label{Fig7}
\end{figure*}

In contrast to experimental observations of the filament radius $a(t)$ (which can be used to infer an evolving stress in the filament given by $\chi/a(t)$), we can directly observe the polymer elongation and stress in our model, and hence better understand the effects of changing the relative concentrations of PS7 and PS16.
In \fref{Fig7} (a) and (b), we plot the effective Weissenberg number $W\!i_\mathrm{eff} = \hat{\tau}_{i1} \dot{\varepsilon}$ and stress contributions $\Delta \sigma_{p,i}$ of the two species in a mixture of 1000ppm PS7 and 500ppm PS16, as per \fref{Fig1} (values of $\tau_{i1}$ can be found in \tref{tab:parameters}).
From these plots, we can clearly observe the three regimes of behaviour.
During the initial IVC regime, polymeric stresses are low as the chains have not yet extended beyond their equilibrium configuration, so it is viscous and inertial forces that balance the capillary pressure during the thinning process. 
The extension rate rapidly increases due to the growing capillary pressure as the filament thins, leading to a high enough Weissenberg number for both species to be extended by the flow (i.e. $\dot{\varepsilon} \hat{\tau}_{i1} = W\!i > 0.5$ for both $i = 1$ and $i=2$). 
This extension rate is fast enough that the elongation is essentially affine (i.e. $\dot{\bm{A}}_{is} \approx \bm{\kappa \cdot \bm{A}_{is}} + \bm{\bm{A}_{is}  \cdot \kappa^T}$), with all chain normal modes for both species being extended equally.
This leads to a jump in the polymer stresses, with greater stress contributed by the more-numerous PS7 species, which has a considerably higher number-average density. 
This can be seen from Eqs.~\ref{eq: polymer modulus} and \ref{eq: polymer stress}, where the same affine stretch $\bm{A}$ leads to a larger stress for the species with the higher number density $n_i$.

This growing polymer stress eventually exceeds the contribution from viscous and capillary stresses, leading to the establishment of an elastocapillary balance.
During this regime, the elongation rate decays, and those modes whose effective Weissenberg number falls below 0.5 will be below the coil-stretch transition and hence their extension $\mathrm{Tr}\bm{A}_{is}$ and contribution to the total polymer stress difference $\Delta \sigma_p$ (given by \eref{eq: polymer stress difference}) will decrease. 
The stress difference in these relaxing modes, which are returning to their equilibrium configuration, decays exponentially, at a rate inversely proportional to the Weissenberg number.
This can be seen more clearly in \fref{Fig7}~(c) and (d), where we have split the molecular weight distribution into $N_j = 30$ individual sub-species (as explained in \aref{Appendix_splitting_MW}) to better visualise the stress evolution of each part of the molecular weight distribution.
The sub-species which just touches $W\!i = 0.5$ sees a constant polymer stress with time during the EC regime, while those above and below see an increasing and decreasing polymer stress respectively.
Were it not for the finite extensibility of the chains, the extension rate would continue decreasing until it asymptotically approached a value such that the highest molecular-weight species had $W\!i = 2/3$ \cite{entov1997effect}. 

In reality, the finite extensibility of the chains leads to an increase in extension rate at long times, as the chains cannot support further elastic stresses by continuing to extend.
As described by Entov and Hinch \cite{entov1997effect}, in this configuration the rate of change of the chain extension $\dot{A}$ becomes negligible, and hence the nonlinear FENE-term becomes proportional to the extension rate, rather than being an exponentially decaying function.
When fully extended, the stress for each species in the terminal regime is given by:
\begin{equation}
\label{eq: stress at max ext}
    \Delta \sigma_{p, is}^\mathrm{terminal} = 2 g_i \tau_{is} L_i^2 \dot{\varepsilon}
\end{equation}
which we note by reference to \eref{eq: force balance} is effectively a visco-capillary balance (stress linearly proportional to strain rate), which generates a linear filament decay and hence breakup in finite time.
This nonlinearity also leads to a minimum in the extension rate at intermediate times, after which the Weissenberg number again begins to climb.

We therefore see that the polydispersity and finite extensibility are intimately coupled when one wishes to use a CaBER experiment to measure effective chain relaxation times.
The measured value of $\tau_\mathrm{EC}$ depends upon the relative proportions of different species within the solution, and upon the strain history experienced by the fluid.
A succinct way to express this idea is that when one measures a decay constant from a CaBER experiment, what one is really measuring is the weighted, conformation-dependent relaxation time of those species which are above the critical coil-stretch transition of $W\!i = 0.5$ during the EC regime (given that one is at a sufficiently high concentration that an elastocapillary timescale can be measured).

\begin{figure}[ht]
    \centerline{
    \begin{tabular}{c}
        \includegraphics[width=8.5cm,height=!]{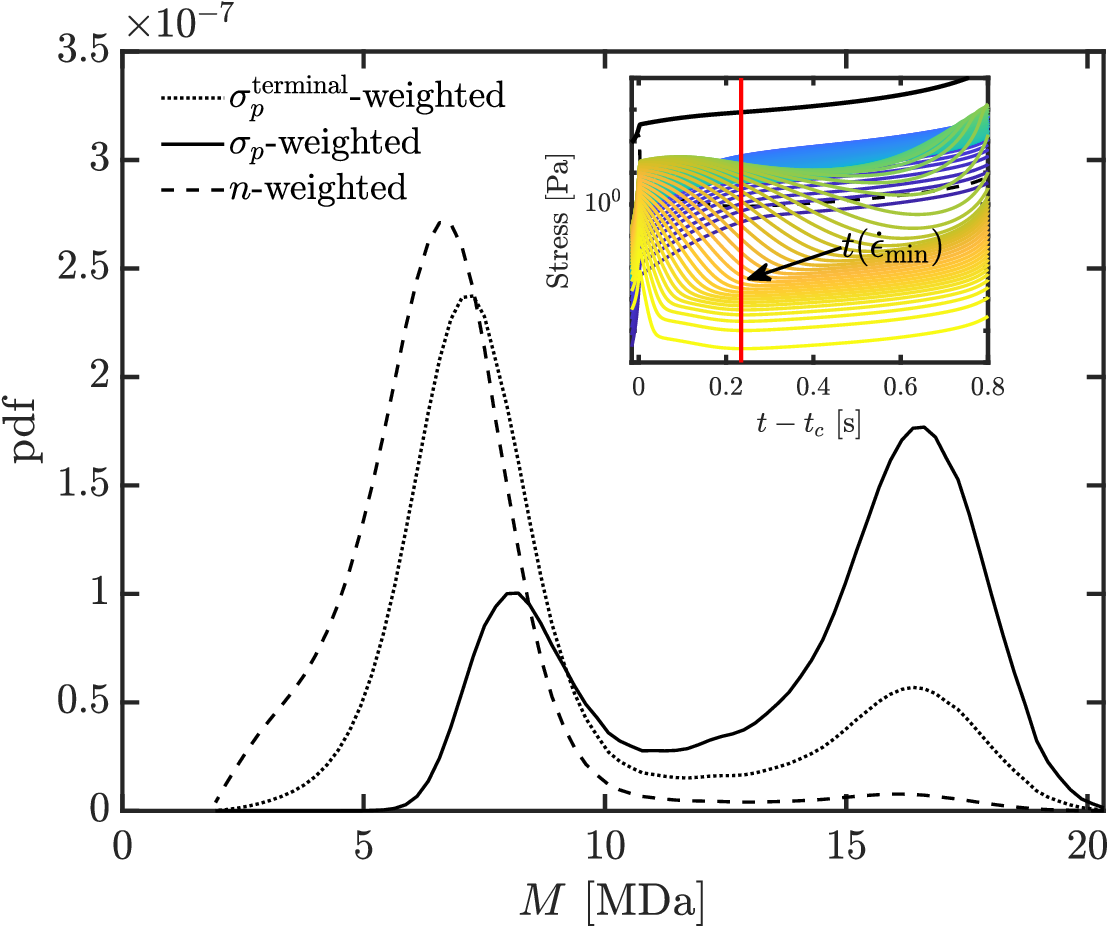} \\
        (a) \\
        \includegraphics[width=8.5cm,height=!]{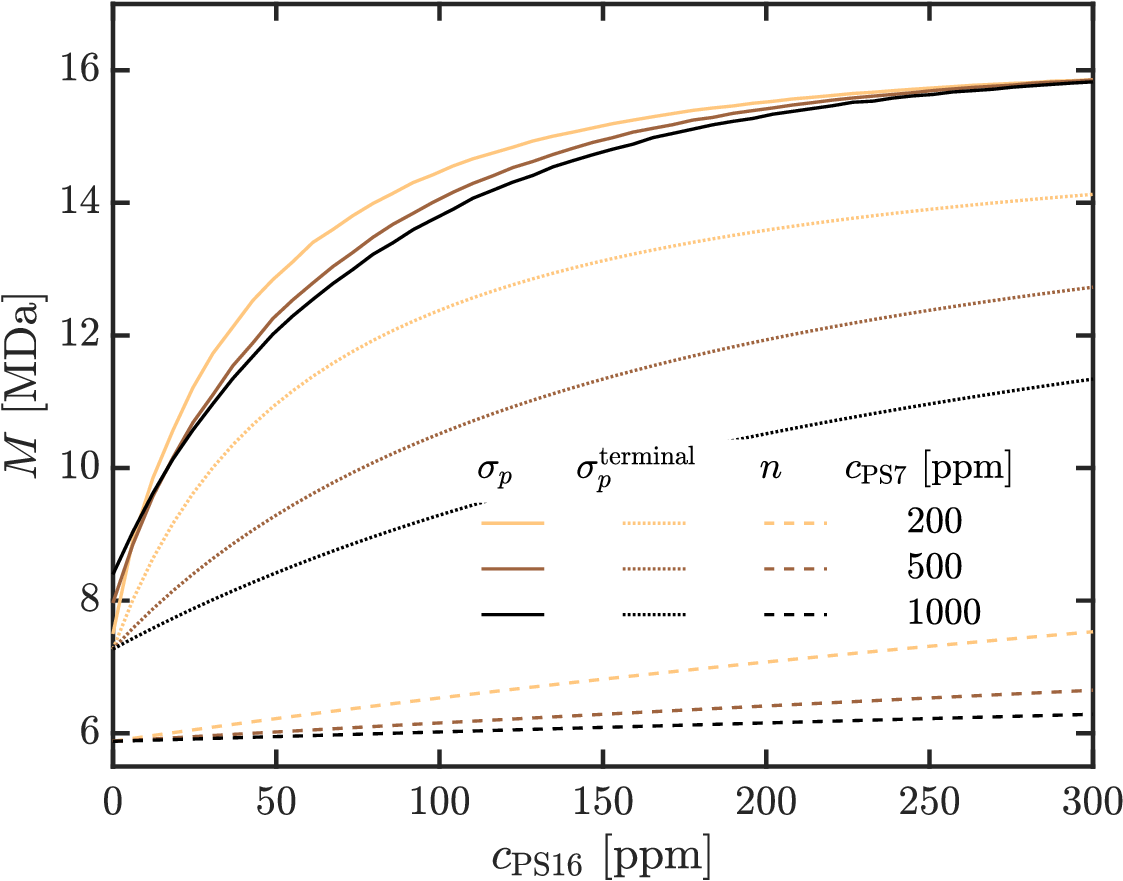} \\
        (b) \\
    \end{tabular}
    }
    \caption{(a) Differently weighted distributions for a mixture of PS7 and PS16. Dashed line is weighted by the number density (directly proportional to the elastic modulus). Dotted line is weighted by \eref{eq: stress at max ext} for each of the $N_j = 80$ sub-species used for the purposes of this plot. Solid line is weighted by the stress contribution at $\dot{\varepsilon}_\mathrm{min}$, as displayed in the inset. (b) The mean molecular weight of each distribution in the top figure for a variety of concentrations of PS7 and PS16.}
    \label{Fig8}
\end{figure}

We quantify this idea in \fref{Fig8}, where we plot differently-weighted probability density functions of the molecular weight distribution.
The number-density weighted average (which is proportional to the elastic modulus of each sub-species $g_i$) is skewed towards the low-molecular weight species, which does not accurately capture their contribution to the elastic stress under elongational flow.
Weighting each species by their stress as in \eref{eq: stress at max ext} and then taking the mean leads to \eref{eq: ext-weighted MW}, the extensibility-weighted average molecular weight.
This distribution is more strongly influenced by the high-molecular-weight, long-relaxation-time, high-extensibility chains, but it assumes that all chains in the ensemble are fully extended.
Finally, we can directly weight the distribution by the simulated contribution to the stress from our model at $\dot{\varepsilon}_\mathrm{min}$, which further `discounts' those chains which are below the coil-stretch transition and hence contribute negligible stress.
From \fref{Fig8}~(b), one can see that the stress-weighted molecular weight correctly implies a considerable dominance of the PS16 species, with the average molecular weight quickly changing from approximately 7MDa to the 16MDa asymptote.
While not exact, the extensibility-weighted molecular weight does predict that high molecular weight species contribute outsized stress.

We note that the molecular-weight dependence of the contributed elastic stress implies that for a highly polydisperse solution, the measured $\tau_\mathrm{EC}$ can grow with $c$ even without explicit concentration effects, as previously discussed by Calabrese et al. \cite{calabrese2025effects}.
With increasing concentration, more elastic stress will be carried by the high molecular weight species, and hence the measured $\tau_\mathrm{EC}$ will be larger. 
Therefore, one should be careful about interpreting findings for $\tau_\mathrm{EC}$ as a function of concentration for extremely polydisperse samples, as it may not imply the occurrence of chain-chain interactions, but rather stretching of different portions of the molecular weight distribution.

\subsection{Pre-stretch and plate diameter}
\label{sec: pre stretch and plate diameter}

\begin{figure}[ht]
    \centerline{
    \begin{tabular}{c}
        \includegraphics[width=8.5cm,height=!]{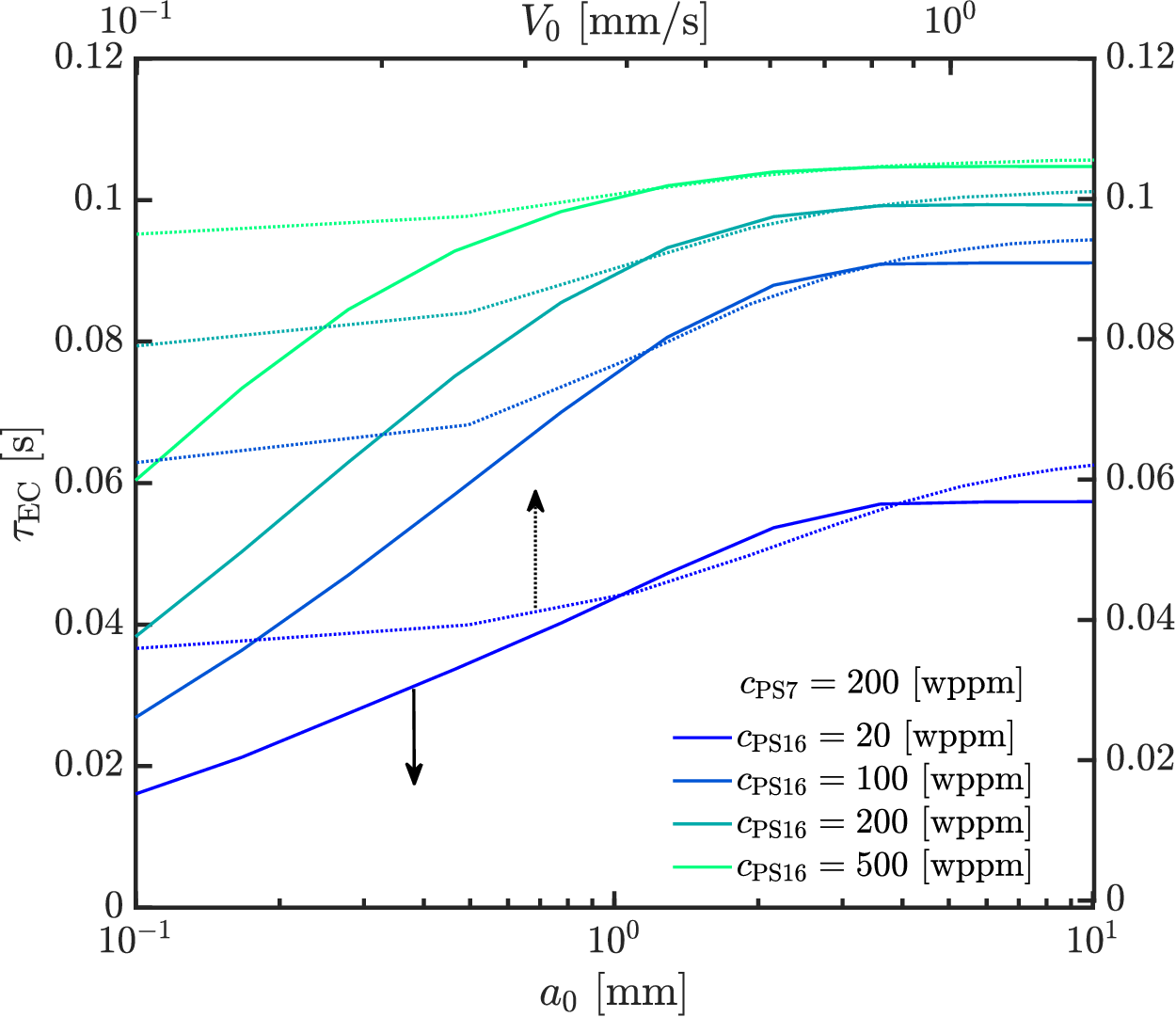} \\
    \end{tabular}
    }
    \caption{Effects of increasing pre-stretch rate and initial filament diameter on the measured $\tau_\mathrm{EC}$ for several concentrations of polymer. Dotted lines correspond to changing the initial pulling speed $V_0$ as per the top axis, while solid lines correspond to changing the initial filament radius $a_0$ as per the bottom axis.}
    \label{Fig9}
\end{figure}

We have seen that for an ideal, linear Oldroyd-B fluid, only the longest relaxation time will be measured in a filament-thinning experiment.
However, for real polymeric fluids, the finite extensibility causes some combination of modes to be stretched, leading to a measured EC timescale which depends upon concentration and polydispersity.
In our simulations dependence also extends to experimental parameters, such as the pre-stretch history and initial filament diameter.
This is shown in \fref{Fig9}, where the measured EC timescale is plotted against both the pre-strain rate (separation rate of the CaBER plates prior to necking in the liquid filament, see \aref{Appendix_pre_stretch}), as well as the initial filament diameter at the point of necking.
Interestingly, both an increase in pre-stretch as well as the plate diameter lead to a longer measured EC timescale, which plateaus at large values of the pre-strain rate or $a_0$.
This is less pronounced at higher concentrations of polymer, which may be why the effect is not seen in some previous studies \cite{miller2009effect}.
Significant pre-stretch means the EC regime is reached at a larger filament diameter, giving more time during the filament decay for the shorter modes to decay away and hence a longer EC timescale to be measured.
Similarly, a larger initial filament diameter gives more time for the shorter modes to decay away in the EC regime.
Both methods converge to similar (although not identical) final values of $\tau_{EC}$ as a function of PS16 concentration.
Overall, it is clear that our modelling supports the conclusion, seen in several recent papers \cite{gaillard2024beware, gaillard2025does, hu2025revealing}, that the measured $\tau_\mathrm{EC}$ in a CaBER experiment is not simply the longest relaxation time $\tau_Z$, but instead also a function of the chain extensibility, fluid viscosity, initial filament diameter, and polymer pre-stretch.

\section{Conclusions}
\label{Conclusions}

This study demonstrates that CaBER experiments on polydisperse polymer solutions measure an effective timescale that is determined by the subset of chains contributing significant elastic stress during the elastocapillary regime. 
Specifically, only chains with Weissenberg numbers above the coil-stretch transition ($W\!i > 0.5$) contribute meaningfully to the polymer stress, and since relaxation time scales as $M^{1.5}$, high molecular weight species maintain this condition longer and thus dominate the measured $\tau_\mathrm{EC}$.
Therefore, our extended FENE-PM model successfully captures the key experimental observation that small concentrations of high molecular weight polymers can dramatically influence filament thinning dynamics \cite{calabrese2025effects}.
The extensibility-weighted molecular weight given by \eref{eq: ext-weighted MW} captures part of this effect, particularly during the terminal regime when all chains are nearly fully extended, but does not account for the dynamic extension of chains during thinning.
Additionally, we show that the measured $\tau_\mathrm{EC}$ is also dependent on experimental parameters such as initial filament diameter or pre-stretch history, in line with recent findings \cite{gaillard2024beware, gaillard2025does}.

These results have practical implications for industrial processing and effective modelling of polydisperse polymer solutions.
The disproportionate influence of high molecular weight species suggests that careful control of the molecular weight distribution tail is crucial for achieving desired flow properties. 
Understanding these features of polydisperse solutions is essential for optimizing industrial processes involving complex polymer mixtures, as well as understanding droplet breakup in disease transmission, or even the spinning of biological fibers.

Future work will focus on incorporating more sophisticated descriptions of chain-chain interactions, in particular self-concentration effects for polydisperse solutions \cite{prabhakar2017effect}.
This could also involve full simulations of the initial IVC kinematics \cite{zinelis2024fluid}, as well as exact Brownian dynamics simulations of the underlying model to remove the approximations associated with FENE-PM averaging \cite{wedgewood1991finitely}. 

\begin{acknowledgments}

We are indebted to Prof. Prabhakar Ranganathan for providing an implementation of the conformation-dependent drag model and helpful comments on its use.
SJH gratefully acknowledges the support of the Okinawa Institute of Science and Technology Graduate University (OIST) with subsidy from the Cabinet Office, Government of Japan, and also funding from the Japan Society for the Promotion of Science (JSPS, Grant No. 24K07332 and 24K00810).

\end{acknowledgments}

\section*{Data Availability Statement}

All data and Matlab scripts used to generate the figures in this paper are available online at \url{https://github.com/IsaacPincus/polydisperseCapillaryBreakup}.
They were originally run in Matlab 2023b, and require the curve fitting toolbox.

\appendix

\section{Second method for splitting distribution}
\label{Appendix_splitting_MW}

\begin{figure}[ht]
    \centering
    \includegraphics[width=8.5cm,height=!]{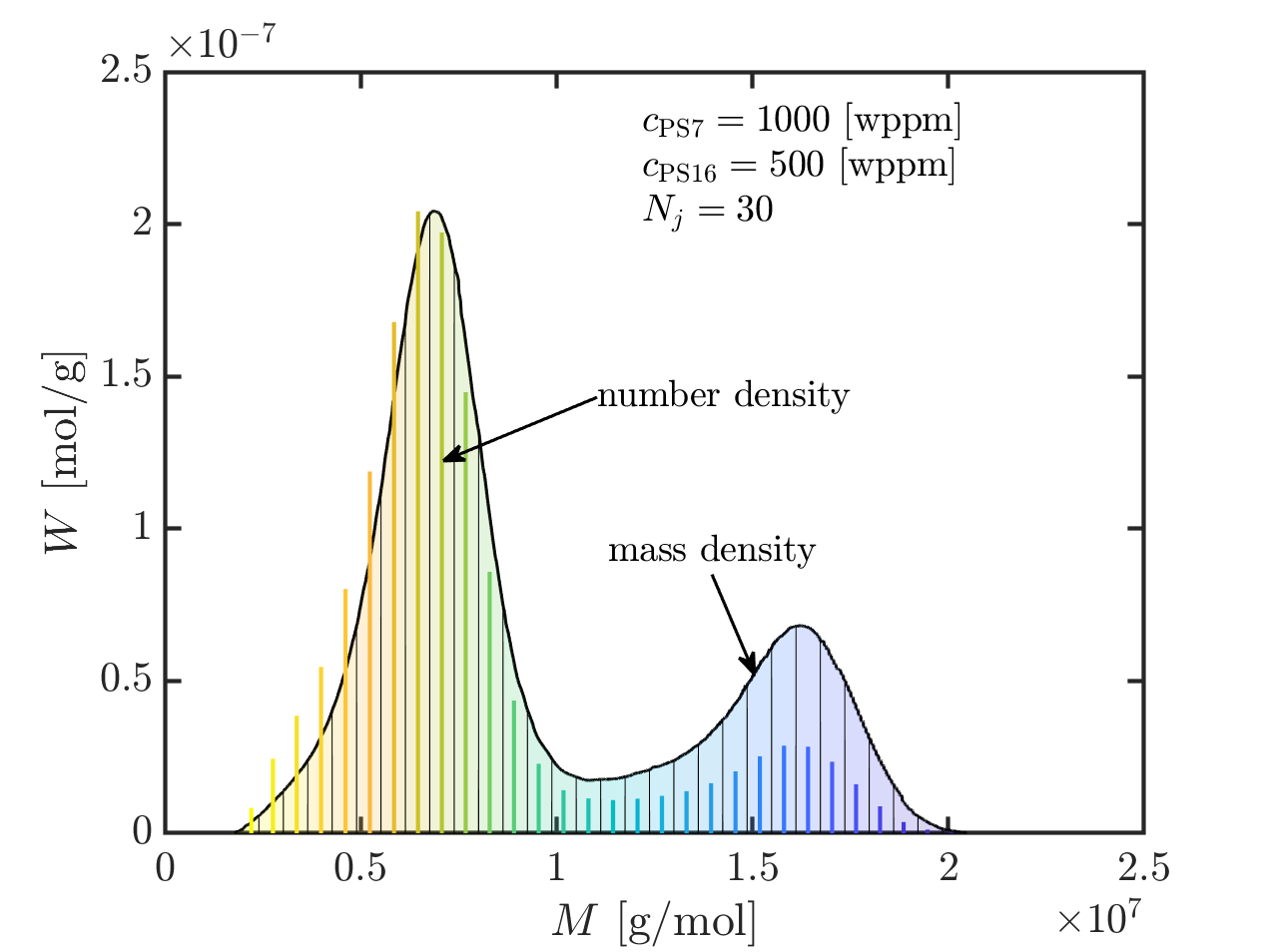}
    \caption{Combined molecular weight distribution of the PS7 and PS16 samples \cite{calabrese2025effects} for 1000 ppm of PS7 and 500 ppm of PS16. The combined distribution is divided into $N_j=30$ bins, each of which represents a separate `sub-species' in our model for Figs.~\ref{Fig7} and \ref{Fig8}. For each bin, there is a line detailing the extensibility-averaged molecular weight for that bin, with the height of the line proportional to the number-fraction of that sub-species in the final ensemble.}
    \label{FigA1}
\end{figure}

For Figs.~\ref{Fig7} and \ref{Fig8}, we split the molecular weight distribution into many individual sub-species.
To do so, we combine the original overlapping PS7 and PS16 molecular weight distributions into an overall distribution $W(M)$:
\begin{equation}
    W(M) = \sum_i^{2} \frac{c_i}{c} W_i(M) \equiv \sum_i^{2} w_i W_i(M)
\end{equation}
where
\begin{equation}
    c = \sum_i^{2} c_i
\end{equation}
and $i$ refers to either the PS7 or PS16 sample respectively.
We then split the distribution back into $N_j$ equally-spaced bins, as shown in \fref{FigA1}.
We reimagine the polymer solution as a mixture of $N_j$ new non-interacting monodisperse sub-species, with the molecular weight of each sub-species given by:
\begin{equation}
    M_{L,j} = \left( \frac{\int_{M \in \mathrm{bin}_j} W(M) M^{1 + \nu} dM}{\int_{M \in \mathrm{bin}_j} W(M) dM} \right)^\frac{1}{1 + \nu}
\end{equation}
which is identical to \eref{eq: continuous ext-weighted MW} except that the integrals now run only over the portion of $M$ which is in the $j$th bin ($M \in \mathrm{bin}_j$).
Similarly, the number-density of each sub-species is then:
\begin{equation}
    n_j = \frac{c \cdot \rho \cdot N_A}{10^6} \int_{M \in \mathrm{bin}_j} \frac{W(M)}{M}\mathrm{d}M
\end{equation}
where again the integral runs only over the $j$th bin.
This is displayed graphically in \fref{FigA1}.

\section{Pre-stretch}
\label{Appendix_pre_stretch}

Here we describe a lubrication model for the purely viscous thinning of the filament during initial plate separation, following the derivation of Spiegelberg et al. \cite{spiegelberg1996role}.
By disregarding the influence of surface tension and applying a lubrication approximation, one arrives at an analogy of the classical viscous squeeze film, only reversed.
This leads to the following set of dimensionless differential equations for the fluid motion:
\begin{align}
    0 = & - \Lambda_0 \frac{\partial p}{\partial \xi}  + \frac{\partial^2 u}{\partial \zeta^2} \\
    0 = & -  \frac{\partial p}{\partial \zeta}  + \frac{\partial^2 v}{\partial \zeta^2}
\end{align}
where $\Lambda_0$ is the initial plate aspect ratio $L_0/R_0$, $L_0$ is the initial plate separation, $R_0$ is the initial plate radius (the fluid is assumed to be a cylinder at $t = 0$), $p$ is the dimensionless pressure (non-dimensionalised using $\eta_s V_0/R_0$), $V_0$ is the initial upper plate velocity, $u$ and $v$ are the dimensionless radial and axial velocities respectively, $\xi$ is the dimensionless radius, and $\zeta$ is the dimensionless axial distance from the bottom plate. 
The time dependence arises through the boundary conditions, with a stationary bottom plate and a top plate with a displacement $H(t)$, where $H(t) = L(t)/L_0$ is the dimensionless displacement, and time is non-dimensionalised using $L_0/V_0$.
This leads to the solution:
\begin{subequations}
\begin{equation}
u(\xi, \zeta, t)=\frac{-3 \xi}{\Lambda_0^2} \frac{\dot{H}}{H}\left(\frac{\zeta}{H}\right)\left[1-\left(\frac{\zeta}{H}\right)\right]
\end{equation}
\begin{equation}
v(\xi, \zeta, t)=\dot{H}\left(\frac{\zeta}{H}\right)^2\left[3-2\left(\frac{\zeta}{H}\right)\right]
\end{equation}
\end{subequations}
We note a misprint in the original paper, which has $H$ rather than $\dot{H}$ in the term for $v$.
From here, we switch to a Lagrangian description of the individual fluid points, which amounts to solving the following differential equations for the positions of the Lagrangian labels $\left(\xi(t; \xi_0, \zeta_0),  \zeta(t; \xi_0, \zeta_0) \right)$ which begin at a position $\left(\xi_0, \zeta_0 \right)$ with $0 \le\xi_0 \le 1, 0 \le\zeta_0 \le 1$:
\begin{subequations}
\begin{equation}
    \dot{\zeta} = \dot{H}\left(\frac{\zeta}{H}\right)^2\left[3-2\left(\frac{\zeta}{H}\right)\right]
\end{equation}
\begin{equation}
    \dot{\xi} = \frac{-3 \xi}{\Lambda_0^2} \frac{\dot{H}}{H}\left(\frac{\zeta}{H}\right)\left[1-\left(\frac{\zeta}{H}\right)\right]
\end{equation}
\end{subequations}
The first equation can be solved using the substitution $\zeta = y H$.
This leads to a separable equation:
\begin{equation}
    \dot{y} H = \dot{H} \left[ -2 y^3 + 3 y^2 - y \right]
\end{equation}
Integrating both sides gives:
\begin{equation}
    \zeta = \frac{H}{2} \left[ 1 \pm \sqrt{1 + \frac{4 z_0}{H - 4 z_0}} \right]
\end{equation}
where the parameter $z_0$ is product of the distance to the bottom plate $(\zeta_0)$ by the distance to the top plate, divided by the square of the distance to the midplane, namely
\begin{equation}
    z_0 = \frac{(\zeta_0 - 1) \zeta_0}{(1 - 2 \zeta_0)^2}
\end{equation}
which goes to zero at the top and bottom plates and negative infinity at the midplane, implying that $\zeta(t; \xi_0, \zeta_0 = 0.5) = H/2$.
Substituting this into the equation for $\dot{\xi}$ and simplifying gives:
\begin{equation}
    \frac{\dot{\xi}}{\xi} = \frac{3}{4 \Lambda_0^2} \frac{\dot{H}}{H} \frac{4 z_0}{H - 4 z_0}
\end{equation}
which has the solution:
\begin{equation}
    \xi = \left( \frac{1 - 4 z_0/H}{1 - 4 z_0} \right)^{\frac{3}{4 \Lambda_0^2}}
\end{equation}
The (dimensionless) extension rate is then given by the fractional rate of decrease of the filament radius (as in \eref{eq: filament radius evolution} in the main paper), namely:
\begin{equation}
    \dot{\varepsilon} = \frac{-2}{\xi}\frac{d \xi}{dt} = \frac{-3}{2 \Lambda_0^2} \frac{4 z_0}{H - 4z_0} \frac{\dot{H}}{H}
\end{equation}
Now when $\zeta_0 \rightarrow 0.5$ (i.e. for material points initially near the filament midpoint),  we find $z_0 = -\infty$, meaning that:
\begin{equation}
\label{eq: pre stretch ep}
    \dot{\varepsilon}_\mathrm{max} = \frac{3}{2 \Lambda_0^2} \frac{\dot{H}}{H}
\end{equation}
and we note that while expressed here in dimensionless form, this final equation is identical if all quantities are dimensional.
Importantly, this correctly predicts that the extension rate of a material element near the midplane is constant for exponential plate separation, where $\dot{H} \propto H$.
For a constant pulling speed, this model predicts a logarithmic increase in Hencky strain with time.
This analysis omits both surface tension and inertial effects, which will play a role at moderate values of the capillary number and Reynolds number respectively.
For sufficiently slow pulling speeds (capillary number $Ca \ll 1$), surface tension will dominate viscous forces, and the filament shape is more accurately described by an analysis which minimises surface area \cite{gaudet1996extensional}.
For sufficiently fast pulling speeds (Reynolds number $Re \sim 1$), fluid inertia will also affect the filament shape.

\bibliography{__references}

\end{document}